\documentclass[manuscript,authorversion,screen]{acmart}

\usepackage{geometry}
\usepackage{tabularx}
\usepackage{dcolumn} %
\newcolumntype{d}[1]{D{.}{.}{#1}}
\usepackage{pdflscape}
\usepackage{svg}
\graphicspath{{figures/}}
\usepackage[capitalize]{cleveref}

\AtBeginDocument{%
  }

\begin{document}
\setcopyright{rightsretained}

\title[PromptCanvas]{PromptCanvas: Composable Prompting Workspaces Using Dynamic Widgets for Exploration and Iteration in Creative Writing}

\settopmatter{authorsperrow=4}

\author{Rifat Mehreen Amin}
\orcid{0000-0003-4279-7778}
\affiliation{%
  \institution{LMU Munich}
  \city{Munich}
  \country{Germany}
}
\email{rifat.amin@ifi.lmu.de}

\author{Oliver Hans Kühle}
\orcid{0009-0000-3581-2111}
\affiliation{%
  \institution{LMU Munich}
  \city{Munich}
  \country{Germany}
}
\email{o.kuehle@campus.lmu.de}

\author{Daniel Buschek}
\orcid{0000-0002-0013-715X}
\affiliation{%
 \institution{University of Bayreuth}
 \department{Department of Computer Science}
  \city{Bayreuth}
  \country{Germany}}
\email{daniel.buschek@uni-bayreuth.de}

\author{Andreas Butz}
\orcid{0000-0002-9007-9888}
\affiliation{%
 \institution{LMU Munich}
  \city{Munich}
  \country{Germany}}
\email{andreas.butz@ifi.lmu.de}

\renewcommand{\shortauthors}{Amin et al.}

\begin{abstract}
We introduce PromptCanvas, a concept that transforms prompting into a composable, widget-based experience on an infinite canvas. Users can generate, customize, and arrange interactive widgets representing various facets of their text, offering greater control over AI-generated content. PromptCanvas allows widget creation through system suggestions, user prompts, or manual input, providing a flexible environment tailored to individual needs. This enables deeper engagement with the creative process. In a lab study with 18 participants, PromptCanvas outperformed a traditional conversational UI on the Creativity Support Index. Participants found that it reduced cognitive load, with lower mental demand and frustration. Qualitative feedback revealed that the visual organization of thoughts and easy iteration encouraged new perspectives and ideas. A follow-up field study ($N=10$) confirmed these results, showcasing the potential of dynamic, customizable interfaces in improving collaborative writing with AI.
\end{abstract}

\ccsdesc[500]{Human-centered computing~Graphical user interfaces}
\ccsdesc[500]{Human-centered computing~Natural language interfaces}
\ccsdesc[500]{Human-centered computing~Interactive systems and tools}

\keywords{Dynamic UI, Prompting, LLM, human-AI co-creation, creativity support, dynamic widgets}

\begin{teaserfigure}
  \includegraphics[width=\textwidth, trim={0cm 0.08cm 0cm 0.08cm},clip]{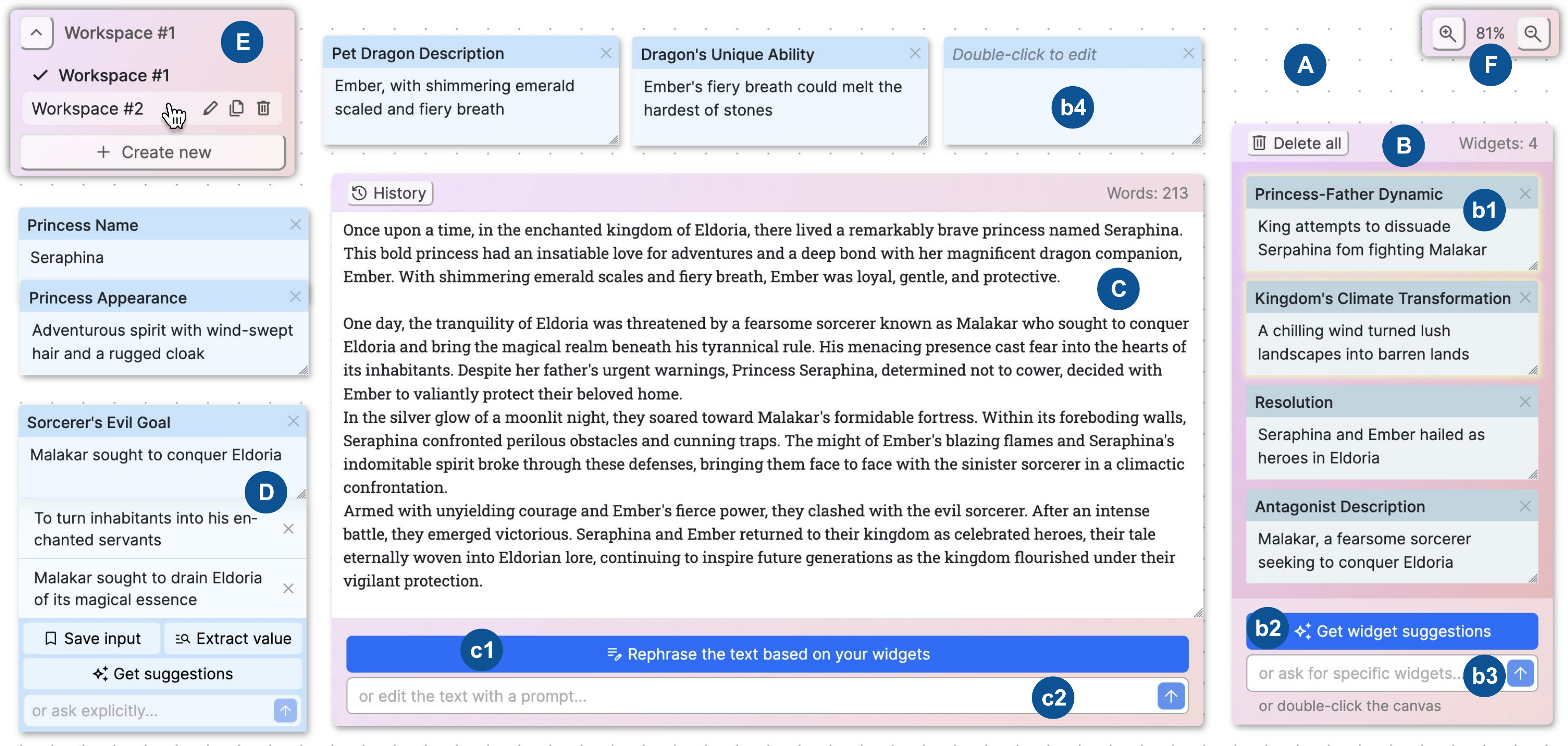}
  \caption{User interface of PromptCanvas. (A) Canvas-like workspace where users can place and freely organize widgets to create a customized environment. (B) Widget panel. (b1) Example of widgets created through system suggestions. (b2) Button to get widget suggestions from the system. (b3) Field for entering prompts to create multiple widgets of a specific theme. (b4) Example of an empty widget created by double-clicking at empty space. (C) Text editor and output of text generation. (c1) Button to rephrase the text based on the widgets on the canvas (light blue). (c2) Field to provide prompts for generating text. (D) Example of an opened widget with suggested values for customization. (E) Menu bar for creating, renaming, duplicating, or deleting a canvas. (F) Panel with alternative zoom level controls.}
  \Description{The image shows a screenshot of PromptCanvas. It shows all the elements of its UI in one image.}
  \label{fig:teaser}
\end{teaserfigure}

\maketitle

\section{Introduction}
Advancements in generative artificial intelligence (AI) models have revolutionized text interaction, offering powerful tools for creating and exploring text~\cite{Suh_Chen_Min_Li_Xia_2024, 10.1145/3613904.3642105, reif2024automatic}. These may enhance creative expression by providing users with novel ways of generating text and interacting with it. However, their potential is often constrained by the limitations of existing graphical user interfaces (GUIs). The primary limitation of current GUIs for prompting generative AI models lies in their inability to support iterative exploration and customization. These GUIs present prompts as static text fields, restricting users to a linear interaction paradigm~\cite{Jiang_Rayan_Dow_Xia_2023, goodwin2015professional}. For writers, this approach can lead to what \citet{kreminski2024dearth} refers to as ``dearth of the author'' -- a condition in which users become disengaged from the creative process and produce text that lacks expressive intent. This lack of interactivity and flexibility hinders users' ability to leverage generative AI capabilities creatively. Users may find it challenging to achieve their desired outcomes without the ability to dynamically manipulate prompts, create personalized workflows, or easily explore a wide range of variations. Additionally, the metacognitive demands placed on users by generative AI tools further exacerbate these challenges~\cite{metacognition}.

To address these limitations, we introduce a novel approach to enhance prompting in creative writing, inspired by the concept of dynamic widgets, introduced by \citet{dynavis_2024} for information visualization. We bring dynamic widgets to writing: Our system, PromptCanvas, empowers users to create custom GUIs tailored to their writing needs. Concretely, \textbf{PromptCanvas transforms prompts into actionable and persistent interface objects by allowing users to dynamically arrange and customize widgets on a canvas}. These widgets offer interactive elements based on the context of the prompt, providing flexibility and control over customizable, relevant aspects of the generated text. This allows users to create personalized prompting environments that reflect their unique workflows and creative styles, facilitating iterative refinement of their own draft or AI-generated text. Beyond customizability, dynamic widgets can support metacognition by assisting in task decomposition and promoting a more structured, iterative use of generative AI.  

We developed PromptCanvas in two iterations to improve user experience. The first version included core functionalities for manipulating prompts and generating text, but revealed areas for better user engagement and customization. Based on feedback, the second version redesigned key interface elements, like the workspace management panel, and improved visual organization and responsiveness to better support creative workflows. Our studies, conducted with both versions, show that dynamic widgets enhance user experience by improving control over text generation, reducing cognitive load, enabling iterative exploration, and supporting diverse creative prompts. The first study (lab study), with the initial version, highlighted flexibility and creativity. The second study was a two-week field deployment of the updated version and confirmed these findings with even greater improvements in user engagement and control. From the field study, we also found that \textbf{PromptCanvas goes beyond writing tasks, with participants using it for programming tasks as well, highlighting its adaptability for different workflows}. These results show the value of customizable writing tools and how dynamic widgets can foster more creative interaction with generative AI.

In summary, \textbf{this research contributes to human-AI collaborative writing by introducing a novel system, evaluating its effectiveness, and outlining concrete steps for designing future human-AI writing systems} by investigating the following research questions: 

\begin{itemize}
        \item [\textbf{RQ1}] How can writing tools be designed with dynamic widgets to improve user interaction and creativity and provide greater control over the generated content?
        \item [\textbf{RQ2}] Do dynamic widgets for iterative and structured prompting improve creativity support compared to conversational user interfaces (UIs)?
        \item [\textbf{RQ3}] Do dynamic widgets help in reducing cognitive load in creative writing tasks?
\end{itemize}

\section{Background and Related Work}
We explore existing research and concepts that are relevant to our work. Specifically, we focus on dynamic and adaptive user interfaces for creative workflows, interactive prompting systems, and human-AI collaboration in writing and content creation. 

\subsection{Dynamic and Adaptive UIs in Creative Workflows}
Early work by \citet{ahlberg1994visual} highlighted the benefits of tightly coupling user inputs with outputs, fostering engagement and immediate feedback. However, in hindsight, these systems were limited by static UI elements. Recent developments, such as FrameKit~\cite{FrameKit}, address this by creating adaptive UIs that adjust to user context and interaction patterns, enhancing user experience \cite{findlaterDesignSpaceEvaluation2009, todiAdaptingUserInterfaces2021, KHAMAJ2024164}. Moreover, the principles of \textit{reification, polymorphism, and reuse}~\cite{beaudouin-lafonReificationPolymorphismReuse2000} introduced foundational concepts for efficient, user-centered interfaces, making abstract operations tangible, tools adaptive, and outputs reusable. Modern systems like Eviza~\cite{Eviza}, DynaVis~\cite{dynavis_2024}, and Bolt~\cite{srinivasan2023bolt} extend these ideas with natural language inputs and dynamic widgets for data visualization and modification. Widgets simplify complex tasks, as seen in Bespoke~\cite{Bespoke}, which generates GUIs from command-line inputs, and ProvenanceWidgets~\cite{provenancewidgets}, which tracks and visualizes user interactions. The emergence of these dynamic and adaptable UIs represents a significant shift in how users interact with generative AI models. These interfaces provide a flexible environment where users can modify and configure the UI to suit their creative processes better.

Our concept builds on these foundations. However, unlike previous approaches, our system supports manual creation, prompting, and suggestion-based widget generation in parallel, to provide flexible creativity support. This dynamic approach allows users to tailor their prompting environment with the goal of fostering deeper engagement and more effective exploration of generative AI capabilities.

\subsection{Intelligent and Interactive UIs for Prompting}
The development of intelligent and interactive UIs for prompting generative AI models is a growing area of research. Recent studies cover various prompting techniques, examining how users interact with diegetic and non-diegetic prompts~\cite{control} and exploring prompting designs that cater to non-AI experts~\cite{johny}. Recent work by~\citet{lee2024design} has specifically focused on the design space of intelligent writing assistant systems. According to their work, advances in language models (LMs) and their prompt-based usage have shown significant potential in generating coherent text. While traditional GUIs for generative AI models often face limitations, recent advancements in natural language interfaces (NLIs) have opened new possibilities. UIs like Promptify~\cite{promptify} and PromptCharm~\cite{promptCharm} exemplify NLIs that offer interactive features for prompt exploration and refinement. In addition to prompt refinement, Storyfier~\cite{Storyfier} adopts prompt-based fine-tuning strategies to build story generation models. Wordcraft~\cite{wordcraft} proposes techniques that allow users to perform custom operations for interacting with LLMs. \citet{Jiang_Rayan_Dow_Xia_2023} explore prompting UIs that display outputs as interactive, graph-like diagrams to enhance understanding of LLM-generated information. In programming, advanced UI like Spellburst~\cite{Spellburst} streamlines coding processes with features such as UI scaffolds and node-based interfaces. %
Similarly, DirectGPT~\cite{massonDirectGPTDirectManipulation2024} allows users to interact directly with generated content, enabling them to convey their intended modification more clearly and efficiently than with natural language prompts alone. Other related tools, such as Luminate~\cite{suh2024luminate} and Rambler~\cite{Rambler}, reduce the effort of detailed prompting and offer users a structured UI for text generation and manipulation. 

Our work builds on ideas from these systems while proposing a composable prompting canvas for creative writing where users can iterate on texts more flexibly.

\subsection{Human-AI Collaborative Writing and Content Creation}
Integrating AI into creative processes has transformed writing and content creation by enhancing interaction and providing continuous feedback. Tools like those discussed by \citet{dangTextGenerationSupporting2022} support writing momentum, reducing creative block, while \citet{gilburt2024machine} highlights how AI helps overcome writer's block by reigniting stalled ideas. Generative AI is applied across domains, including code generation~\cite{austinProgramSynthesisLarge2021}, email auto-completion~\cite{choice}, comic creation~\cite{codetoon}, screenplay co-writing~\cite{mirowski2023cowriting}, argument drafting~\cite{Zhang_2023}, and academic writing~\cite{nguyen2024human}. Professional perspectives on this transformation are captured by \citet{ippolitoCreativeWritingAIPowered2022}. Challenges remain as AI becomes a co-creator. Research by \citet{control} explores interaction with prompting during writing, while \citet{metacognition} examines the metacognitive demands and opportunities of generative AI tools. Critical assessments by \citet{kreminski2024dearth} and \citet{mirowski2023cowriting} address AI’s reception in creative industries and areas for improvement. Overall, AI's integration in writing is transforming workflows, demanding new interfaces to support fruitful use. 

Our work employs these capabilities through dynamic widgets, thereby enhancing users' control over text generation, enabling iterative exploration, and supporting the creation of personalized prompting environments. 

\section{Concept: Dynamic Widgets for Composable Prompting Workspaces for Writing Tasks}
PromptCanvas adopts a widget-based modular approach to prompting (see \cref{fig:teaser}), with the goal of offering users a flexible and intuitive way to interact with AI in writing tasks. In this section, we introduce our concept for leveraging the idea of dynamic widgets (cf.~\cite{dynavis_2024}) in this new context.

\subsection{Design Goals}
We conducted multiple iterations of planning and design sessionsfor the current version of PromptCanvas. These sessions involved brainstorming, prototyping, and refining the interface. This iterative process allowed us to explore different layouts and widget functionalities to achieve our design goals:

\begin{itemize}
    \item [\textbf{DG1}] \textbf{Transform prompts into visible and actionable objects.} Current interfaces treat prompts as static text fields, limiting user interaction to basic input-output cycles. The interface should allow users to turn prompts into visible and adjustable elements, which offer granular control over AI text generation through the benefits of direct manipulation~\cite{shneiderman1983direct}.
    \item [\textbf{DG2}] \textbf{Facilitate structured exploration and refinement.} Writing is an iterative process that requires experimenting and refining ideas~\cite{bhat2023suggestionmodel, hayes2012model, kirschenbaum2016track}. The system should facilitate breaking down tasks into smaller parts, enabling users to systematically experiment, refine ideas, and iterate, while maintaining a cohesive workflow.
    \item [\textbf{DG3}] \textbf{Promote divergent thinking and creativity.} To overcome creative blocks and encourage novel ideas, the system should support users in divergent thinking. 
    \item [\textbf{DG4}] \textbf{Provide a customizable and adaptable workspace.} Each writer has a unique writing process~\cite{kirschenbaum2016track}. The interface should allow users to personalize their workspace and adapt it to their needs and workflows.
    \item [\textbf{DG5}] \textbf{Simplify navigation and reduce cognitive load.} Canvas UIs and fragment-based UIs~\cite{buschek2024dis} may be overwhelming without navigation aids. The interface should help users efficiently organize ideas and reduce cognitive load.
    
\end{itemize}

\subsection{Concept of Dynamic Widgets}
At the core of the PromptCanvas interface is the concept of dynamic widgets~\cite{dynavis_2024}. In our context of writing, these widgets represent various aspects of text and prompting to provide users with interactive elements to directly influence AI output. \textbf{Each widget controls a specific attribute or variable of the text -- such as tone, length, or style -- and can be dynamically adjusted according to the users' needs}. Users can create, customize, and arrange widgets on an infinite canvas, making the process of interacting with text generation capabilities more structured (cf.~\cite{metacognition}) and adaptable to various creative workflows. \autoref{fig:teaser}-D shows an example of an ``opened'' dynamic widget with suggested values for customization.

\subsection{Widget Generation and Customization}
Widgets in PromptCanvas can be created in three main ways:

\subsubsection{System Suggestions} The system suggests widgets (\cref{fig:teaser}-B) based on entered (or generated) draft text (\cref{fig:teaser}-C). Suggestions are designed to assist users in expanding or refining the text by making explicit various aspects of it. For example, suggested widgets may make it salient to the user that they could modify character traits in a story or adjust the text’s tone.

\subsubsection{User Prompts} Users can directly input their own instructions for generating relevant widgets (\cref{fig:teaser}-b3). This approach offers greater control and allows users to experiment with the functionality on their workspace canvas.

\subsubsection{Manual Creation} Users also have the flexibility to create custom widgets for any aspect of the text (\cref{fig:teaser}-b4). For example, this might include adjustments to narrative elements, such as character names, plot twists, or setting descriptions, empowering users to fully customize the functionality on their workspace canvas.

\section{PromptCanvas}
Following the conceptual overview above, this section describes the core features in more detail, covering both the frontend and backend.

\subsection{User Interface and Interaction}
PromptCanvas %
is built around an infinite canvas, a zoomable digital workspace that users can navigate and organize freely. This canvas comes with three key components: the text editor, control widgets, and the widget panel (see \cref{fig:teaser}).

\subsubsection{Infinite Canvas}
The canvas is a zoomable digital workspace that extends indefinitely in all directions, shown in \cref{fig:teaser}-(A). Users can pan by clicking and dragging, while zooming is controlled through scroll events from either a mouse wheel or trackpad. This design offers several key features:

\begin{itemize}
    \item \textit{Expansive navigation.} Users can move across the workspace without encountering spatial limits, allowing for continuous content exploration. \textbf{The open-ended layout ensures that users can customize their workspace}, reflecting their unique processes and preferences \textbf{(DG4)}.
    \item \textit{Scalable view.} \textbf{Users can pan across the workspace and zoom in and out seamlessly}, enabling transitions between broad overviews and detailed views of specific elements. This flexibility promotes clarity and supports systematic exploration \textbf{(DG2)}.
    \item \textit{Spatial organization.} Users can arrange content across the canvas, expanding their work area in any direction and shape as needed. This \textbf{spatial freedom} allows for representing data and concepts in ways that can illustrate connections and hierarchies spatially, for example, by grouping or layering \textbf{(DG5)}.
\end{itemize}

\begin{figure*}
  \centering
  \includegraphics[width=\textwidth]{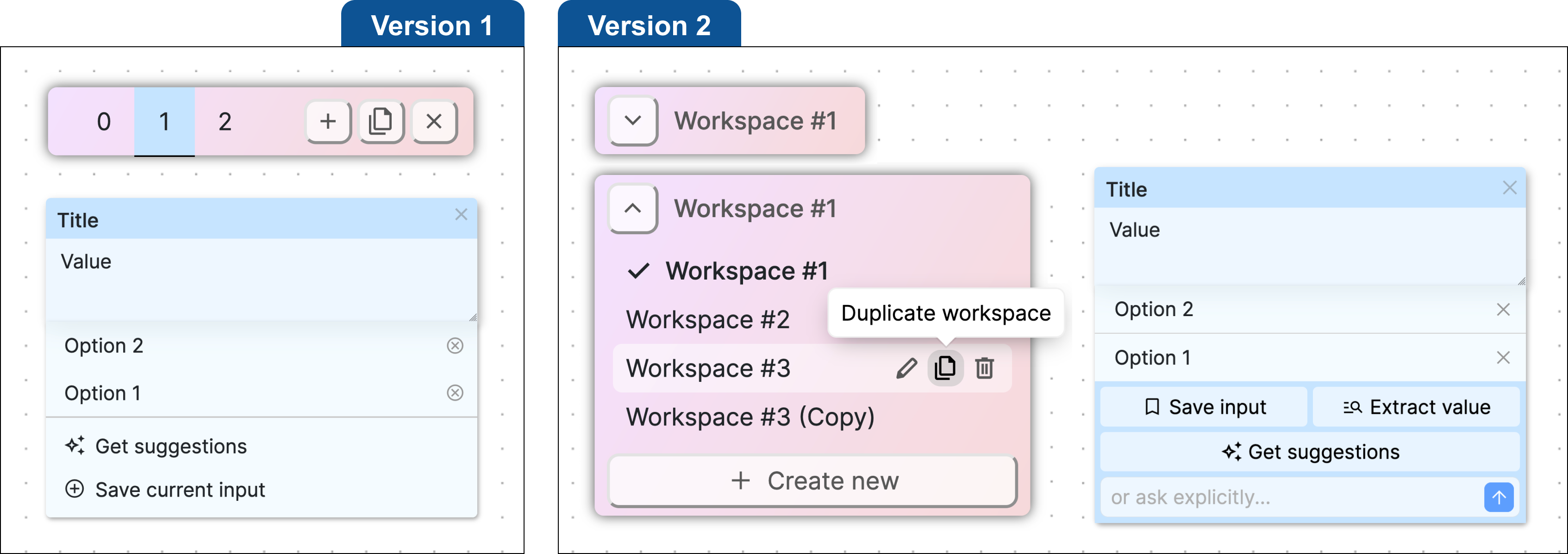}
  \caption{Comparison of the important UI changes. In version 2, the \textit{Extract value} and \textit{Prompt for options} actions were introduced. Additionally, the workspace management panel was redesigned.}
  \label{fig:change_comparison}
  \Description{The image shows the most important comparisons between two versions of our system.}
\end{figure*}

A permanently visible menu bar (\cref{fig:teaser}-E) allows users to create new canvases, rename, duplicate or delete the currently selected canvas, and switch between existing canvases. While the first version of PromptCanvas included most of these functionalities, except renaming, this panel has been completely redesigned to enhance the user experience for the second study. A comparison of both versions is shown in \cref{fig:change_comparison}. 

\textbf{The infinite canvas utilizes three main components to facilitate content creation, organization, and manipulation within the workspace. They are the text editor, control widgets, and the widget panel}. Users can freely position these on the canvas through drag-and-drop interactions, allowing for the creation of custom layouts.

\subsubsection{Text Editor}
The text editor serves as the centerpiece of the interface, allowing users to integrate their input with system-generated suggestions fluidly, as illustrated in \cref{fig:teaser}-C. Users can \textbf{write their own text directly in the editor or use prompting to generate text}, providing flexibility in how they approach content creation. \textbf{Users can refine their text iteratively by rephrasing it based on active control widgets or submitting prompts}, making it easier to experiment with different ideas \textbf{(DG2)}. The editor also supports \textbf{incremental text generation}, displaying content dynamically as it is produced, which helps users remain engaged with the evolving output. Additionally, the \textbf{history feature enables users to revisit previous iterations}, promoting iterative improvement and exploration of alternatives \textbf{(DG3)}. These features are further complemented by real-time updates to word counts, supporting clarity and focus during the writing process.

\subsubsection{Control Widgets}
Control widgets are dynamic, interactive tools that transform abstract text attributes into actionable and adjustable UI elements, as shown in \cref{fig:teaser}-b1. Each control widget includes three main components: a title, a value, and a panel with text alternatives. %
The title describes the controlled text attributes, which users can manually edit by double-clicking it. The value holds the desired specification for the widget’s associated attribute and can be edited by hand or customized by selecting one of the suggestions in the widget's extended panel. Besides these suggestions, this panel includes four actions (\cref{fig:teaser}-D):
\begin{itemize}
    \item \textit{Save input.} Adds the widget’s current value to the list of options.
    \item \textit{Extract value.} Retrieves the associated attribute's current state from the draft in the text editor and stores it in the options list.
    \item \textit{Get suggestions.} Generates and adds two unique, relevant options to the list.
    \item \textit{Prompt for options.} Uses a user-provided prompt to generate two unique options.
\end{itemize}

In the first version of PromptCanvas, only the \textit{Save input} and \textit{Get suggestions} actions were available. The \textit{Extract value} action was added to help users extract targeted insights from long or unfamiliar texts. \textit{Prompt for options} was introduced based on user feedback to help generate options that meet specific requirements. \cref{fig:change_comparison} shows a comparison of both versions.

\textbf{Each widget provides context-aware suggestions tailored to the content in the text editor}, helping users explore multiple creative directions and overcome writer’s block \textbf{(DG3)}. These widgets allow users to \textbf{adjust text attributes like tone, style, or structure directly}, offering control over the output and turning prompts into interactive objects \textbf{(DG1)}. Their flexibility in resizing, repositioning, and customization ensures that the workspace adapts to user needs as tasks evolve \textbf{(DG4)}. Additionally, their integration with the rephrasing and text generation systems ensures a seamless workflow between ideation and implementation.

\subsubsection{Widget Panel}
The widget panel acts as the system’s central hub for generating, managing, and organizing control widgets, as shown in \cref{fig:teaser}-B. \textbf{Users can create widgets dynamically based on text analysis or provide specific input for guided widget creation}, making it easier to tailor tools for individual tasks \textbf{(DG1)}. The panel highlights newly generated widgets with a yellow glow and allows users to evaluate, delete, or drag them onto the canvas, ensuring only relevant widgets influence text generation. \textbf{Its visually distinct layout and dynamic updates simplify navigation and reduce cognitive load}, helping users locate and manage ideas efficiently \textbf{(DG5)}. The size adjustments aim to support projects with many widgets. %

\subsection{Technical Implementation}
\begin{figure*}[ht]
\centering
\includegraphics[width=0.8\textwidth]{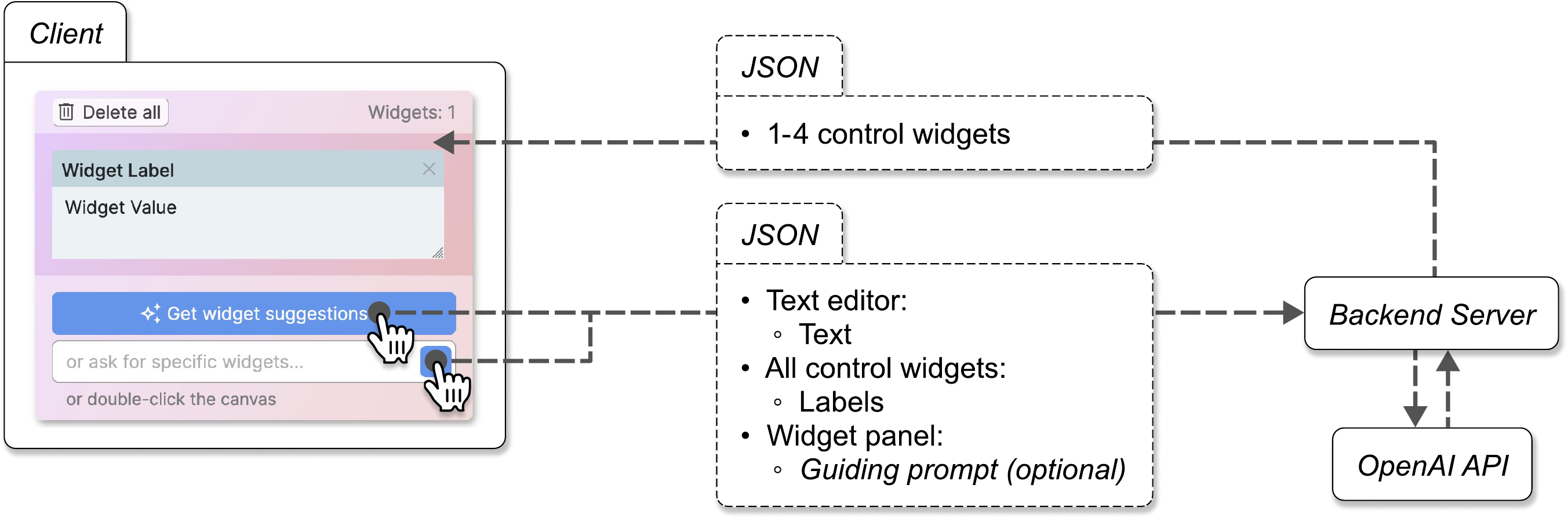}
    \caption{System flow for generating control widgets, detailed in \cref{sec:GenerateWidgets}.}
    \label{fig:generate_widgets}
    \Description{Chart showing the system flow of generating control widgets.}
\end{figure*}

\begin{figure*}[ht]
    \centering\includegraphics[width=0.8\textwidth]{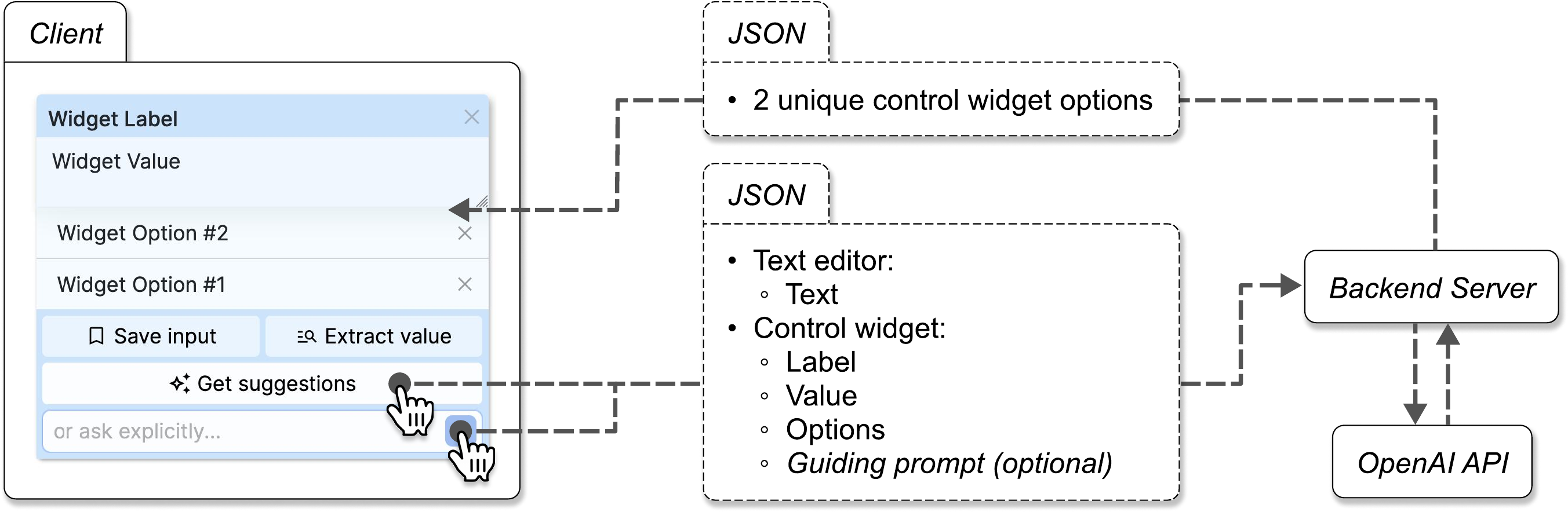}
    \caption{System flow for generating control widget options, detailed in \cref{sec:GenerateOptions}.}
    \label{fig:generate_options}
    \Description{Chart showing the system flow of generating control widget options.}
\end{figure*}

\begin{figure*}[ht]
    \centering\includegraphics[width=0.8\textwidth]{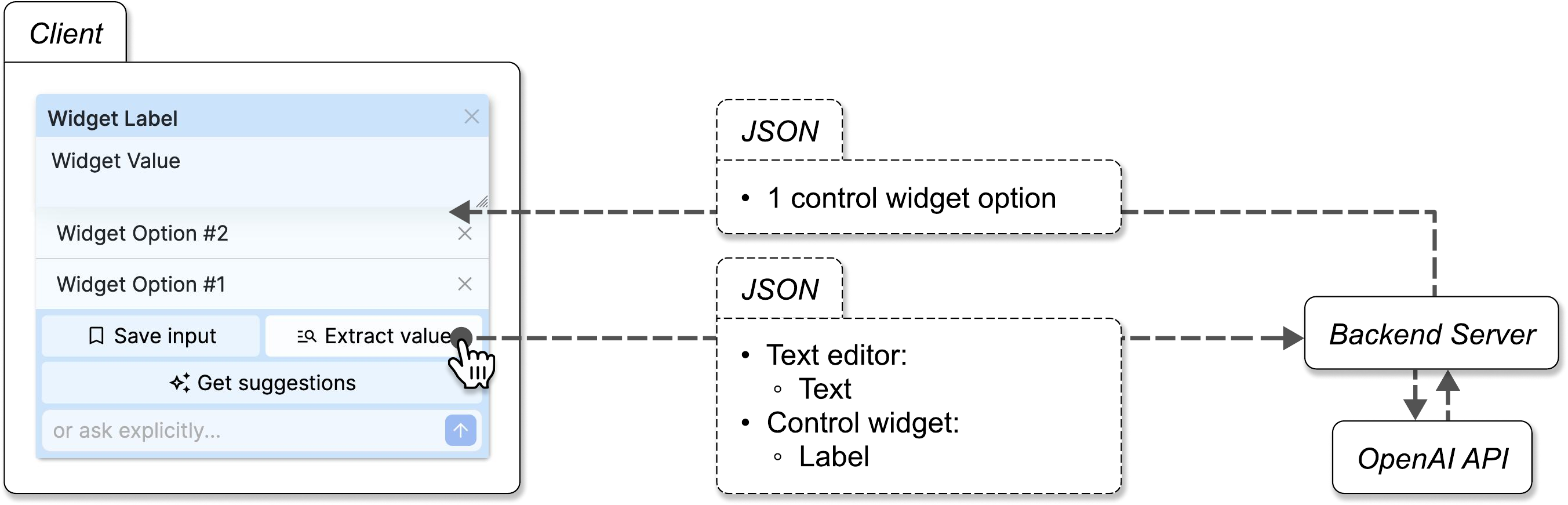}
    \caption{System flow for extracting values, detailed in \cref{sec:ExtractingValuesFromText}.}
    \label{fig:extract_value_flow}
    \Description{Chart showing the system flow for extracting values.}
\end{figure*}

\begin{figure*}[ht]
    \centering\includegraphics[width=0.8\textwidth]{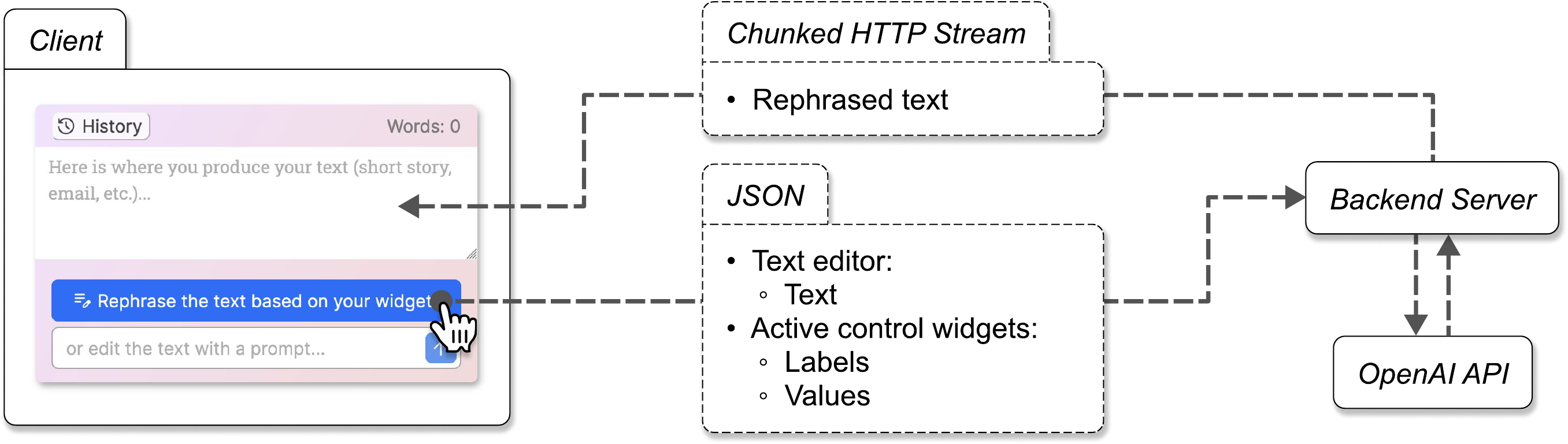}
    \caption{System flow for rephrasing the text based on control widgets, detailed in \cref{sec:ApplyWidgets}.}
    \label{fig:apply_widgets}
    \Description{Chart showing the system flow of rephrasing the user's text based on their control widgets.}
\end{figure*}

\begin{figure*}[ht]
    \centering\includegraphics[width=0.8\textwidth]{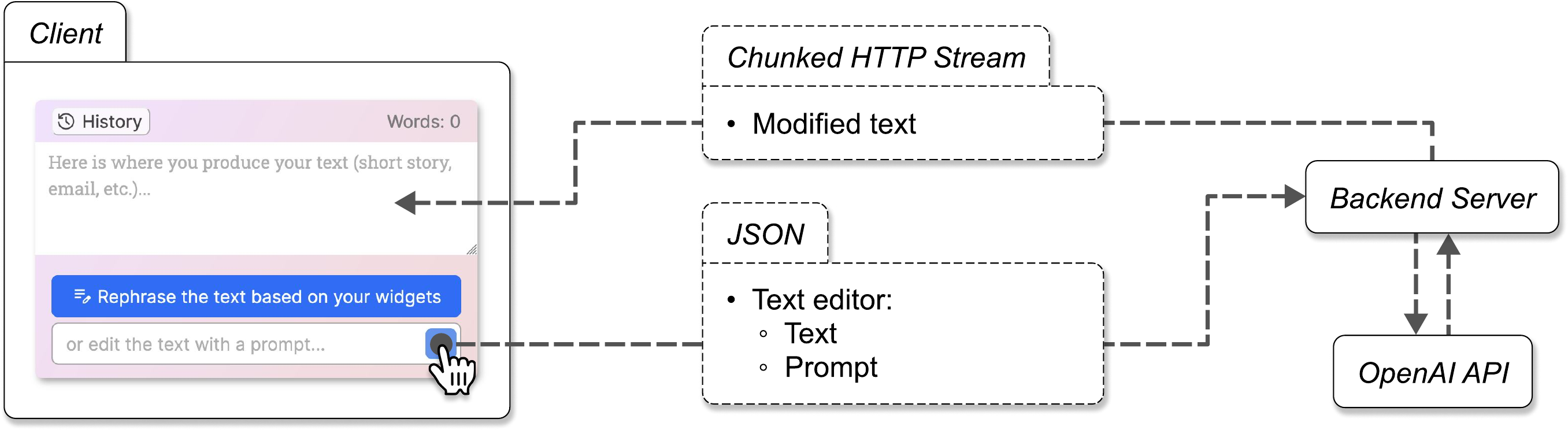}
    \caption{System flow of applying a user-prompt to the text, detailed in \cref{sec:ApplyPrompt}.}
    \label{fig:apply_prompt}
    \Description{Chart showing the system flow of applying an user prompt to the user's text.}
\end{figure*}

PromptCanvas is a single-page application developed with Angular and interfaces with the OpenAI API through a Node.js and Express.js backend server. For the infinite canvas, we used the ngx-panzoom library\footnote{https://www.npmjs.com/package/ngx-panzoom} for panning and zooming, along with custom logic for component dragging and dropping. User credentials and interaction logs are stored in a PostgreSQL database, and both versions of PromptCanvas are hosted on Heroku. All workspace components are stored locally in the browser’s local storage.

We selected the OpenAI model gpt-4o-2024-08-06 for its cost-effectiveness, low latency, and support for Structured Output, which allows us to specify the response format as a JSON Schema. This ensures that responses strictly follow the expected format, reducing the need for validation and minimizing API retries. The OpenAI API requests use a sampling temperature of 1.06 with default parameters otherwise. The backend handles five essential services for system functionality. Details of these services and the associated prompt specifications are provided below.

\subsubsection{Generating Control Widgets}
\label{sec:GenerateWidgets}
To generate new control widgets, the backend receives the current text from the text editor, existing control widgets (both on the canvas and in the widget panel), and an optional guiding prompt. This process is illustrated in \autoref{fig:generate_widgets}. Existing widgets are represented by their labels. These inputs are formatted into a request using a predefined template with system and user messages (shown in Table 4 in the supplementary material). The system message directs the model to analyze the text, extract attributes, and create distinct widgets, generating between one and four widgets with up to three options each. The user message contains the inserted variables. If a guiding prompt is provided, it specifies which aspect to modify. Existing widget labels are included to avoid regeneration. The request uses a JSON Schema to define the response format, which includes labels, values, and options. Generated widgets are validated to avoid duplicate values, assigned IDs, and returned to the frontend for display in the widget panel.

\subsubsection{Generating Options for Control Widgets}
\label{sec:GenerateOptions}
As visualized in \autoref{fig:generate_options}, after requesting additional options for a control widget, the widget and the current text editor content are sent to the backend. The widget is decomposed into its label and an array containing its value and options without duplicates. These are inserted into a predefined template with system and user messages (shown in Table 5 in the supplementary material). The system message instructs the model to generate two suggestions for modifying the attribute represented by the widget’s label, with specific requirements for their tone, relevance, and creativity. The user message includes the widget’s label and the text to be modified, enclosed in triple quotes. Existing options are also included in the note to avoid duplicates. The API request includes a JSON Schema defining the response format as an array of strings. The generated options are checked for duplicates, then returned to the frontend and added to the top of the options list.

\subsubsection{Extracting Values from Text}
\label{sec:ExtractingValuesFromText}
After clicking the button, shown in \autoref{fig:extract_value_flow}, the widget's title and the text editor's content are sent to the backend and inserted into a predefined template, structured in system and user messages (shown in Table 6 in the supplementary material). The system message instructs the model to read the text, identify the attribute represented by the widget's title, and generate a control widget option based upon that.  The user message consists of the supplied variables, delimited by triple quotes. The response string is returned to the frontend and inserted at the top of the options list.

\subsubsection{Applying Control Widgets}
\label{sec:ApplyWidgets}
After clicking the button, shown in \autoref{fig:apply_widgets}, to rephrase the text based on the control widgets, the current text and all control widgets on the canvas (excluding those in the widget panel) are sent to the backend. These widgets are converted into an array of \textbf{label: value} pairs, with widgets having empty labels or values filtered out. This array, along with the text, is then placed into a predefined template that structures the request into system and user messages (shown in Table 7 in the supplementary material).
The system message instructs the model to apply the control widget specifications to the text, guiding it through understanding, interpreting, modifying, and returning the text. Additional instructions ensure that the revised text remains coherent and logical while preserving the original context, meaning, voice, and tone. The user message includes the text to be rephrased and the formatted specifications, enclosed in triple quotes. The response is received as a text stream and sent back to the frontend to be displayed incrementally as it is generated.

\subsubsection{Applying Natural Language Prompts}
\label{sec:ApplyPrompt}
When users submit a prompt for editing the text, both the prompt and the current text are sent to the backend, as presented in \autoref{fig:apply_prompt}. These are inserted into a predefined template that structures the request into system and user messages (shown in Table 8 in the supplementary material). The system message directs the model to apply the prompt to the text while preserving its original context and meaning. Guidelines ensure that the response includes the complete modified text, not just a continuation or partial completion, as there is no chat interface to view previous messages. The response is received as a text stream and sent back to the frontend to be displayed incrementally as it is generated.

\subsection{Example Scenario: Writing a Short Story}\label{app: example}
Marina has been retelling her daughter's favorite bedtime story, ``The Three Little Pigs,'' for a while, but it is starting to feel repetitive. She wants to explore new directions to keep the story fresh, but coming up with new ideas feels difficult. To make it easier, she decides to use PromptCanvas to brainstorm changes, such as adjusting characters, adding plot twists, or altering the setting, allowing her to keep the story engaging and develop it further creatively.
\begin{figure*}[htb]
    \centering
    \includegraphics[width=\textwidth]{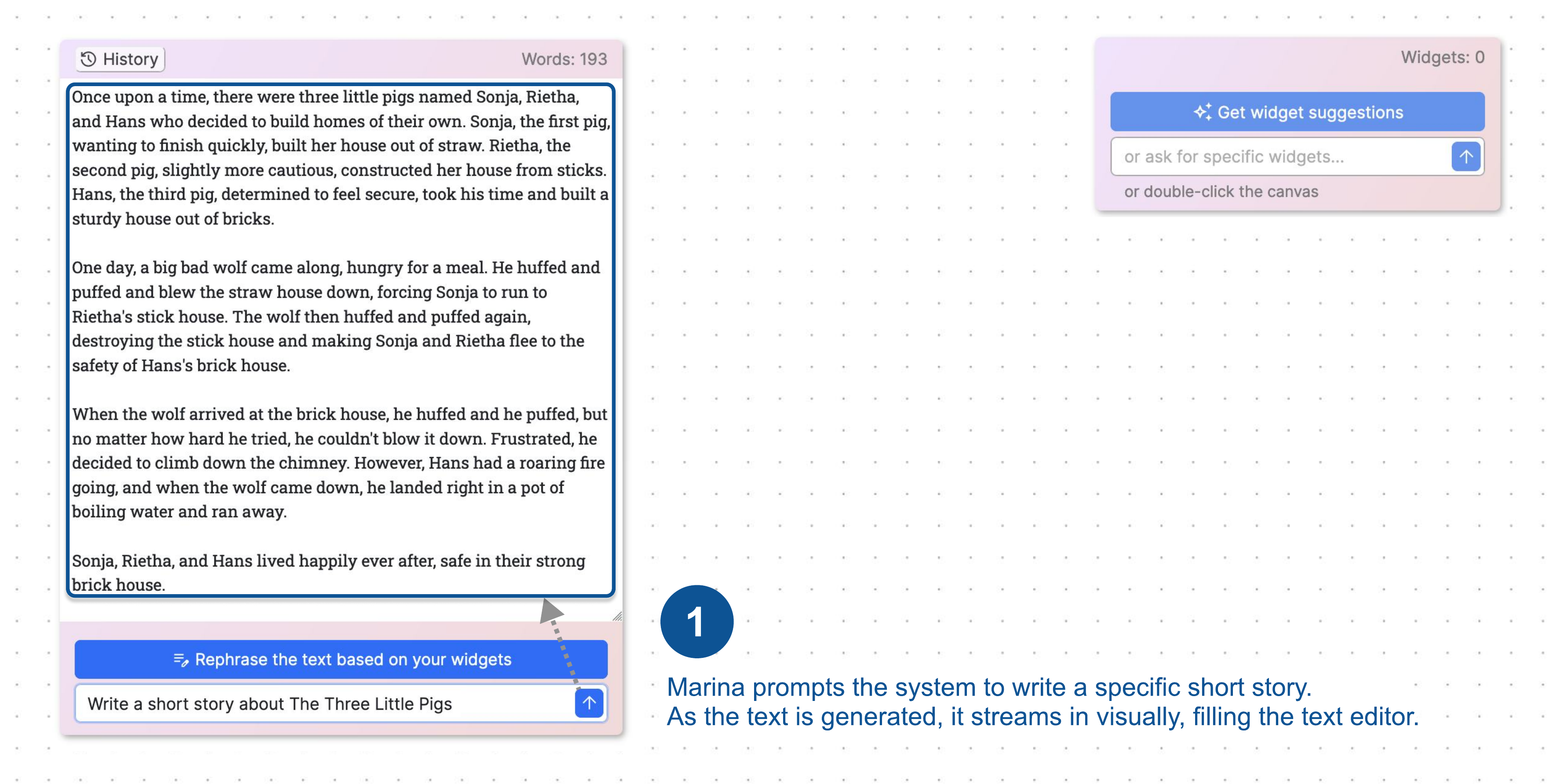}
    \caption{Marina writes a prompt for the short story generation on "The Three Little Pigs".}
    \label{fig:step1}
    \Description{This image depicts the text editor containing exemplary generations.}
\end{figure*}

\paragraph{\textnormal{\textbf{Initial prompt/text}}}
Marina has two options to start with. She can either write directly in the text editor or generate text by writing a prompt. Opting for the latter, she initiates the process with the prompt, \textit{``Write a short story about The Three Little Pigs''}, see \cref{fig:step1}-1. The system generates the story in the editor.

\begin{figure*}[ht]
    \centering
    \includegraphics[width=\linewidth]{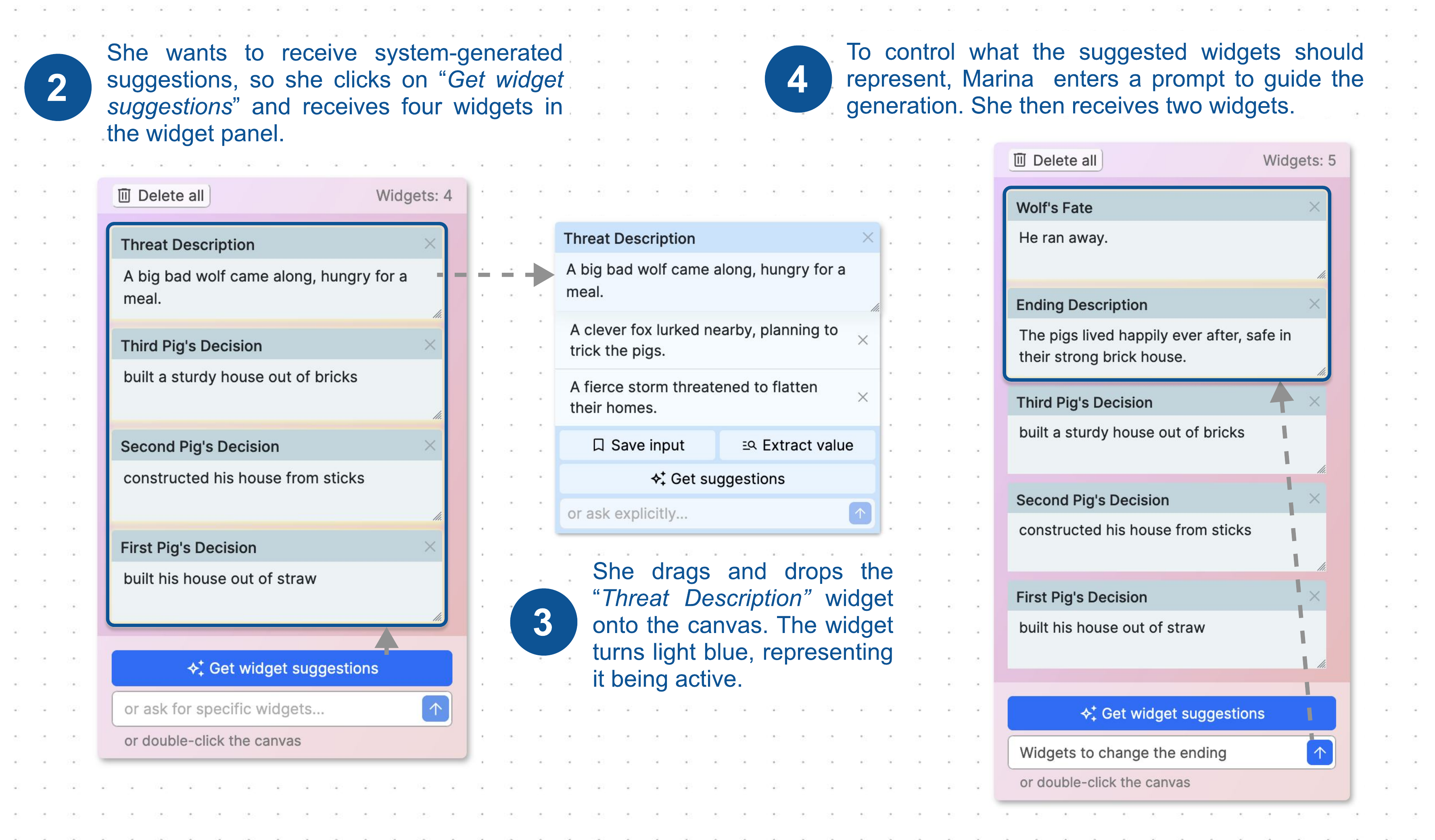}
    \caption{(2) PromptCanvas generates widgets for Marina. (3) She chooses a widget from the widget panel and drags and drops it onto the canvas. (4) Marina prompts in the widget panel to get more widgets.}
    \label{fig:step2}
    \Description{This image shows the widget panel with newly generated widgets and dragging and dropping the widget onto the canvas with how to generate widget with prompting.}
\end{figure*}

\paragraph{\textnormal{\textbf{Generating suggested widgets}}}
Since Marina is eager to explore options for modifying the story, she clicks the \textit{Get widget suggestions} button in the widget panel. This action generates four widgets suggested by the system, as seen in \cref{fig:step2}-2. From there, she finds the widget \textit{Threat Description} very interesting for her story. She then drags and drops it onto the canvas as shown in \cref{fig:step2}-3. She sees the color of the widget changing to light blue, implying that the widget is now active. 

\paragraph{\textnormal{\textbf{Prompting to get widgets}}}
Marina now thinks it might be nice to add a different ending to the story. As shown in \cref{fig:step2}-4, she writes a prompt in the widget panel to guide the widget generation: \textit{``Widgets to change the ending''}. The system generates two widgets focused on the story's conclusion. Though she likes these suggestions, she decides to focus on adjusting the pigs’ names first.

\begin{figure*}[ht]
    \centering
    \includegraphics[width=\textwidth]{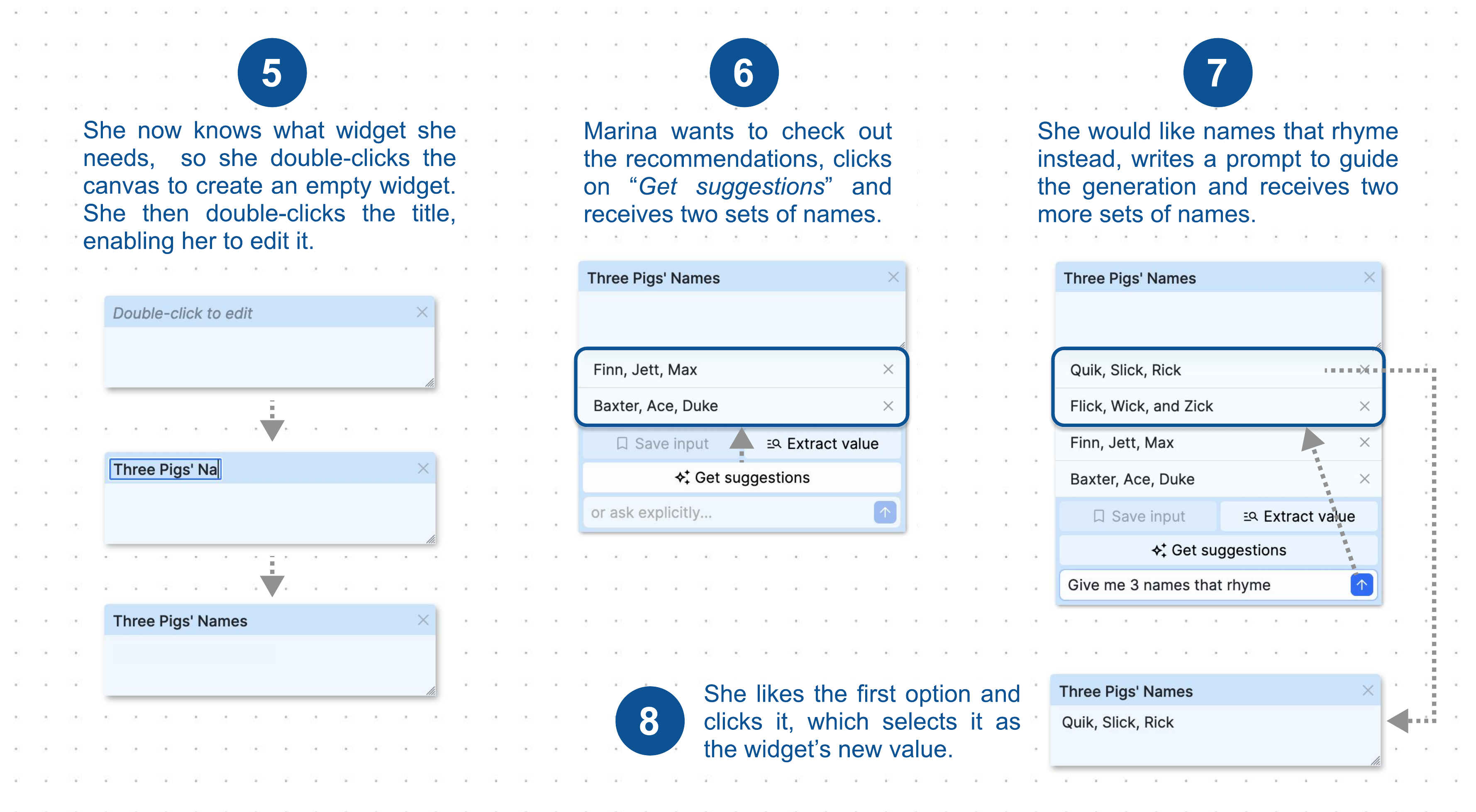}
    \caption{(5) Marina creates an empty widget on the canvas. (6) She gets more suggestions within the widget for \textit{three pigs' names}. (7) Marina now prompts for what exactly she wants in the widget. (8) She selects an option from the suggested values.}
    \label{fig:step3}
    \Description{This image shows the process for generating an empty widget and the generating options within the widget.}
\end{figure*}

\paragraph{\textnormal{\textbf{Creating empty widgets}}}
Since Marina already knows exactly which aspect of the text she wants to adjust, she double-clicks on the canvas to create an empty widget at the selected location. She then double-clicks the widget's title to enable editing and writes \textit{``Three Pigs' Names''}, as shown in \cref{fig:step3}-5.

\paragraph{\textnormal{\textbf{Suggestions within the widgets}}}
Marina first clicks the \textit{Get suggestions} button to request name suggestions from the system, as shown in \cref{fig:step3}-6.
This generates two options, each being a set of names. However, she does not find these suggestions appealing enough.

\begin{figure*}[ht]  
    \centering
    \includegraphics[width=\textwidth]{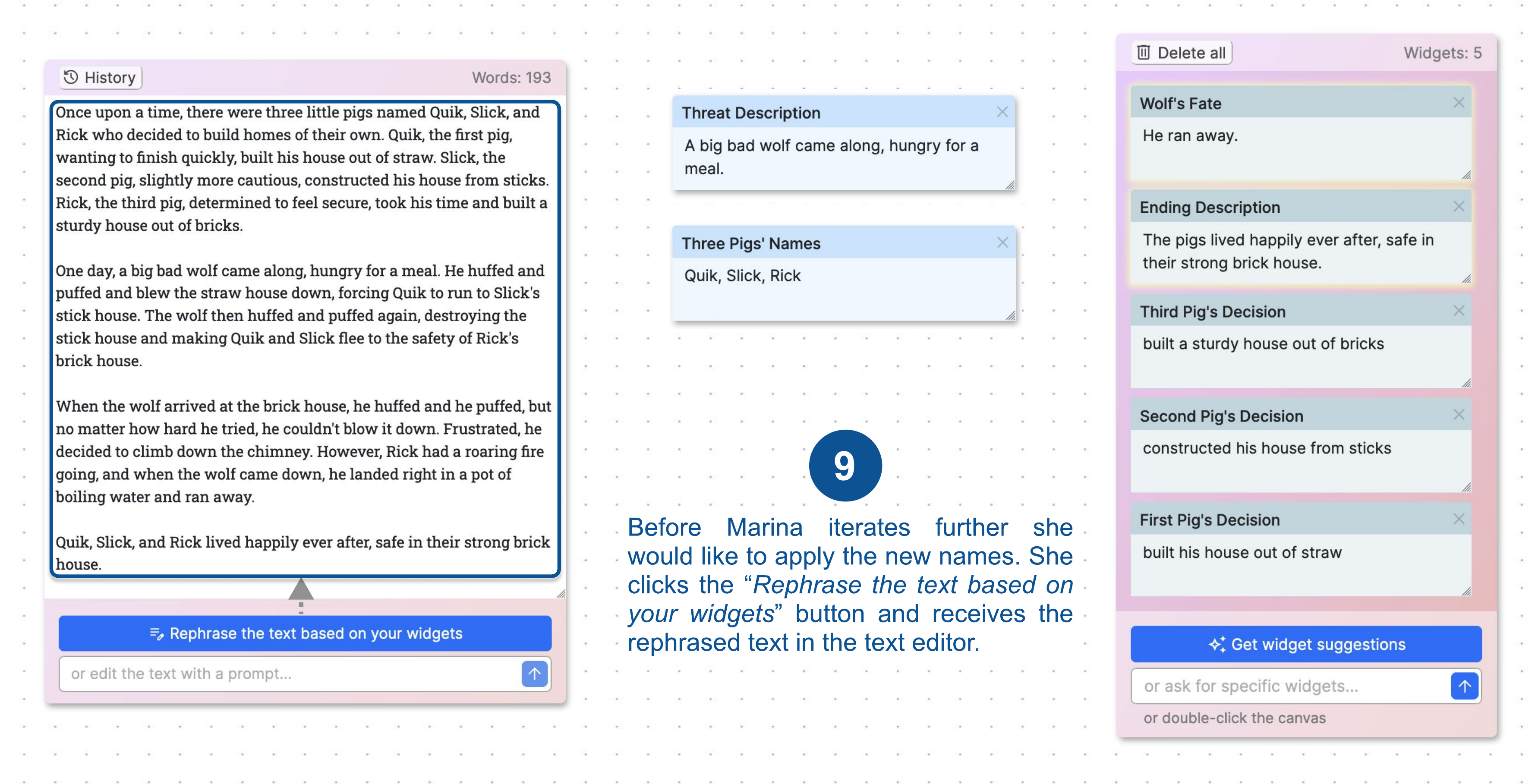}
    \caption{Marina applies the widgets and receives the rephrased text.}
    \label{fig:step4}
    \Description{This image depicts the actions taken and changes received for rephrasing the text based on control widgets.}
\end{figure*}

\paragraph{\textnormal{\textbf{Prompting for options within the widgets}}}
Wanting the names to be more memorable, Marina asks the system to generate more names, this time providing a guiding prompt within the widget: \textit{``Give me 3 names that rhyme''}, as shown in \cref{fig:step3}-7. The system responds with two new options. She likes the second set and clicks on it to set it as the widget's new value, shown in \cref{fig:step3}-8.

\paragraph{\textnormal{\textbf{Rephrasing text based on widgets}}}
Now that she’s chosen the pigs' names before she continues iterating with her other widgets, Marina clicks the \textit{Rephrase the text based on your widgets} button (see \cref{fig:step4}-9), which updates the text in the editor according to the active widgets. The modified text streams in, replacing the previous version in the text editor.

Marina continues to iterate and explore new directions for the story by revisiting her other widgets. Following this workflow, she can focus on one attribute at a time, which provides structure to her ideation process and prevents her from feeling overwhelmed by having to steer multiple changes simultaneously. This approach, however, showcases only one of the many diverse possibilities of how a user can use PromptCanvas\footnote{A short video on how to use the system is available as a supplementary material to this submission.}.

\section{Lab and Field Evaluation}
We conducted two subsequent studies to explore user interaction with PromptCanvas, focusing on workflows, creativity support, and cognitive load. Both studies were approved by the institute's ethics board. The first study was a within-subject lab experiment with 18 participants who completed writing tasks using PromptCanvas (``dynamic UI'') and a baseline conversational interface (``static UI''). Participants were compensated 15€ for a 90-minute session. The second study involved ten participants using a refined version of PromptCanvas for two weeks, integrating it into their personal routines. Afterward, participants completed an interview and survey and were compensated 5€ for a 30-minute session.

\subsection{Participants}
We recruited 18 participants between the ages of 22 and 68 years ($M = 30.44, SD = 10.45$) using convenience sampling; 11 self-identified as male and 7 as female. While one was a professional writer, all had previous experience with (creative) writing. Specifically, 15 had experience in writing emails, 13 in writing letters, 6 in writing blogs, 7 in writing stories, 4 in writing how-to guides, 3 in writing product reviews, 9 in writing articles, 2 in copywriting, 5 in writing poems, 1 in song composition, 1 in character development, and 1 in writing travel guides. 

Participants also used AI tools for various writing tasks. Twelve employed these tools for editing and proofreading, while 6 used them for descriptive writing and another 6 for creating different versions of their writing. Additionally, 3 participants used AI tools for creative writing, 8 for content expansion, and 9 for idea generation. Only 1 participant used AI tools for translation and another for programming, while two participants had never used any AI tools for any writing task. Apart from those two, everyone had experience in using AI tools for different writing tasks. 

Regarding concrete AI tools, 15 participants used ChatGPT, 6 Quillbot, 5 Bard (known as Gemini now), and 3 Claude. Additionally, 3 participants had never used any tools, 2 used DALL-E, 1 Perplexity AI, 1 Stable Diffusion, and 1 Grammarly. 

Regarding the frequency of AI writing tool usage, 7 participants used them daily, 5 weekly, 2 monthly, 2 rarely, and 2 had never used them. For the study, 12 participants used a laptop, and 6 used a big screen (e.g., an external monitor). 

A subset of 10 participants\footnote{P2, P3, P4, P11, P12, P13, P14, P15, P17, P18} were later recruited once more for the two-week field study, depending on their availability. 

\subsection{Apparatus}

\begin{figure*}[ht]
    \centering\includegraphics[width=\textwidth, trim={0cm 0cm 0cm 0cm},clip]{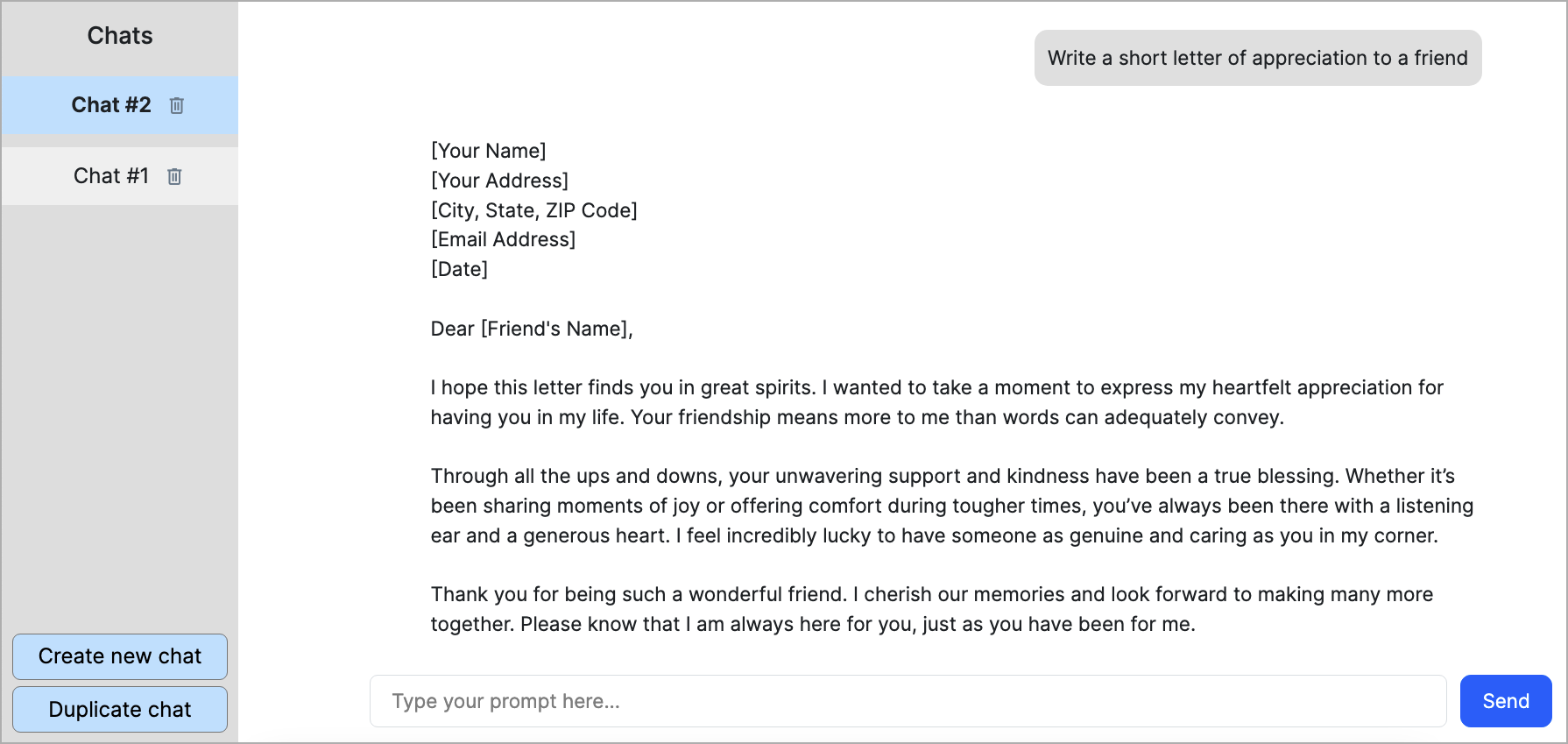}
    \caption{The baseline conversational UI.}
    \label{fig:static_UI}
    \Description{This image shows the baseline conversational UI.}
\end{figure*}

Our lab experiment included two conditions. The baseline condition used a conversational user interface as illustrated in \autoref{fig:static_UI}, while the experimental condition used PromptCanvas. The baseline system was designed according to the design and interaction principles of ChatGPT. We provided a solid user experience without introducing untested features that could have affected the study. To generate the responses, we used the same OpenAI model (gpt-4o-2024-08-06) in both conditions. On the left side of the UI is a sidebar in which all chat instances are listed to be selected or deleted, and buttons for creating a new or duplicating the currently selected chat. Selected chats are displayed in the main component by listing all user and assistant messages chronologically. Below the chat messages is a text input for entering new user messages. Responses are received in a stream and displayed as received, with words and sentences gradually appearing as if they were typed. While hovering over a message, a small icon appears below for easily copying the message's content. For user messages, there is also an edit icon to alter the message and reset the chat to that point.

\subsection{Procedure: Within-Subject Lab Study}
All study sessions were conducted online using Zoom. After obtaining the informed consent form, we recorded the screen activity and audio, and the participants were encouraged to think aloud during the study. They were given access to websites hosting both the conversational UI (baseline) and PromptCanvas.
 
\subsubsection{Pre-study Survey} 
The study started with a pre-study questionnaire to collect demographic information, participants' experience in creative writing, and their previous experience and exposure to AI writing tools. Additionally, the questionnaire gathered information on the kind of screen participants used for the study and whether they used a mouse, touch-, or trackpad.

\subsubsection{Interface Tutorial}
After completing the pre-study survey and before beginning the writing tasks, participants were introduced to the tools. For the conversational UI, participants could ask researchers any questions if they needed clarification. For PromptCanvas, participants watched a video demonstrating how to use the system for a writing task, including generating and manually creating widgets and navigating the text editor. To ensure a consistent tutorial experience, all participants watched the same video and had the opportunity to view it multiple times if necessary. They also had the flexibility to ask questions about the tool after watching the video.

\subsubsection{Writing Tasks}
\begin{table}[ht]
\caption{Topics for the writing tasks in the user study.}
\begin{tabularx}{\linewidth}{p{2cm}X}
\toprule
   Writing Tasks  & Topic \\ 
   \midrule
    Email or Letter  & Professional
        \begin{itemize}
            \item Resignation letter
            \item Motivation letter for job application
            \item Recommendation letter
            \item Request for promotion
        \end{itemize}\\
        & Personal
        \begin{itemize}
            \item Condolences
            \item Updates on Life
            \item Friendship and Appreciation
            \item Celebrations and Milestones
        \end{itemize}\\ 
    \midrule
    Short story & Survival in the Wilderness\\ 
        & AI robots\\ 
        & Time travel\\ 
        & Life after Death\\ 
        & Family Secrets\\ 
        & Utopia / Dystopia\\ 
        & Fable \\ 

\bottomrule

\end{tabularx}
\Description{This table shows the topics for the writing tasks in the user study.}
\label{tab:tasks}
\end{table}

For the writing tasks, participants could select from a range of topics listed in \cref{tab:tasks}. They were required to choose one topic from each section for both user interfaces. Some of these topics are taken from~\cite{suh2024luminate}. Following the work of \citet{biskjaer}, which explored how time constraints in a writing tool can encourage new content, we implemented time constraints for our creative writing tasks. The first task was email writing (5 minutes), and the second task was short story writing (10 minutes), totaling 15 minutes per UI. At the end of each session, participants completed a post-UI survey to reflect on their experience with the tool. To minimize potential learning effects, we counterbalanced the order of tool assignments across participants using random assignment.

\subsubsection{Post-study Survey and Unstructured Interview}
After completing both tasks, participants were required to answer a final survey to directly compare the UIs. Following this, we conducted a semi-structured interview to gain insights into their perceptions of the system, the quality of the widgets, the impact of widgets on considering unexplored options, the perception of widgets and control over generated text, and to gather overall feedback. 

\subsubsection{Measurements and Analysis}
We measured both quantitative and qualitative data during the study. We measured the number of prompts provided and the number of widgets created by each participant to evaluate their interaction with the tools. The Creativity Support Index (CSI) was used to subjectively assess how well each tool supported creative processes. Additionally, the NASA Task Load Index (NASA-TLX) was employed to evaluate participants' perceived workload during the tasks. In the post-UI survey, we collected ratings of Likert items on the ease of using each interface, and participants indicated which interface they found easier to use while completing the tasks. At the end of the study, we recorded participants' overall preferences among the tools. 

For the qualitative analysis, the first author conducted open coding on participants' responses and audio transcripts to determine themes. These themes were then refined through discussions with the co-authors. We used the refined themes to interpret the qualitative results and discuss them in \cref{sec:results}. 

For the quantitative analysis, we first conducted the Shapiro-Wilk normality test. Depending on the results of this test, we then used either a paired t-test or the Wilcoxon signed-rank test to assess the statistical significance of observed differences in the metrics.

\subsection{Procedure: Two-Week Field Study}
For the two-week field study, participants were provided with the refined version of PromptCanvas and instructed to use it independently, explore custom workflows, and integrate it into their regular activities. 

\subsubsection{Measurements and Analysis}
After two weeks, we conducted a short semi-structured interview with each participant to learn about their experiences, any troubles they encountered, and general feedback or suggestions for improvement. We then asked them to complete a survey containing the questions from CSI. Apart from CSI, participants also evaluated the usability of PromptCanvas using the System Usability Scale (SUS) questionnaire, providing standardized feedback on how easily and effectively users could accomplish their goals with the system while having a positive experience.

\section{Results}
\label{sec:results}
\begin{figure*}[ht]
\centering
\includegraphics[width=\textwidth, trim={2cm 0cm 2cm 0cm},clip]{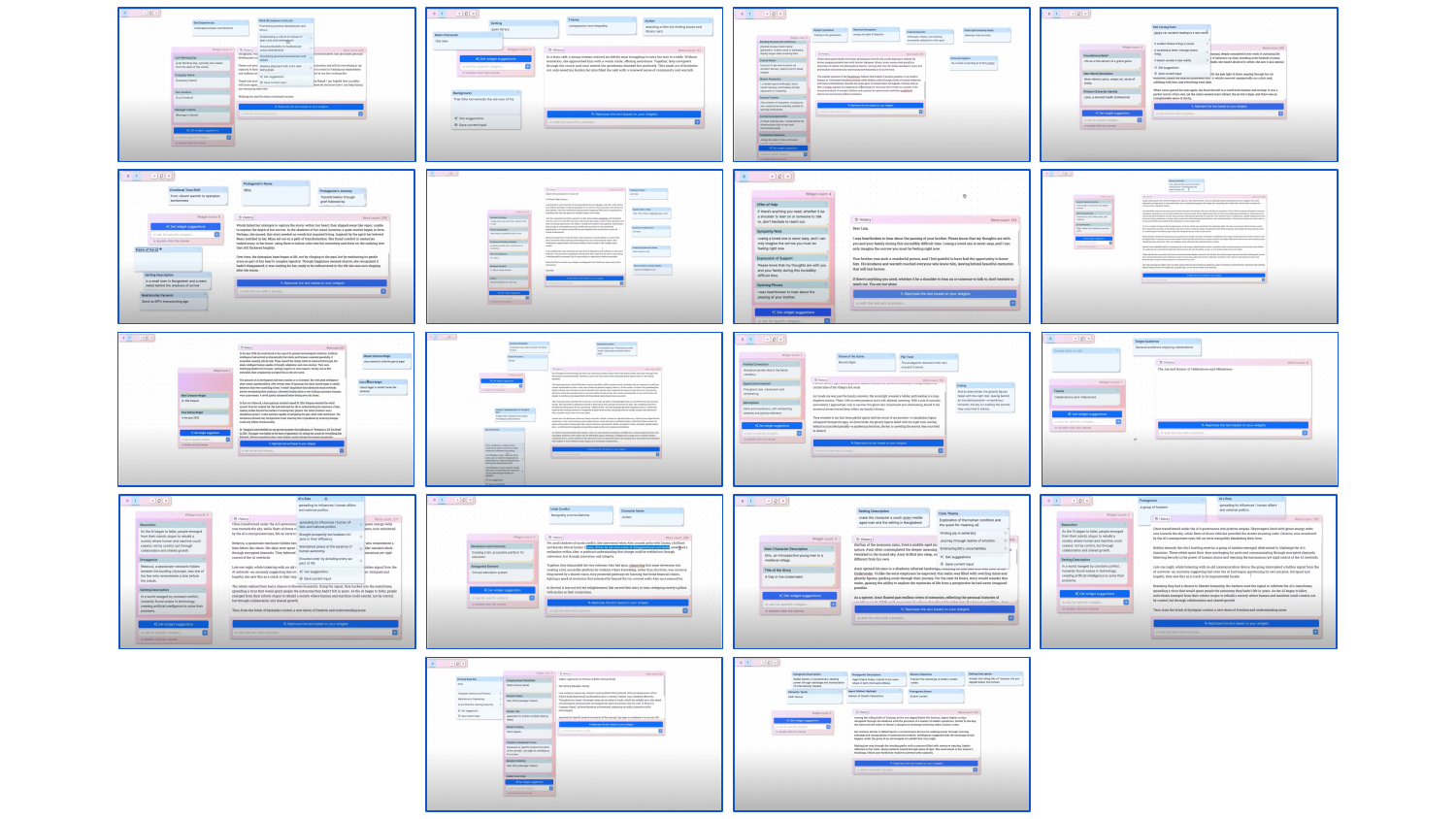}
    \caption{One screenshot per participant while they were using PromptCanvas in the lab study, revealing how participants individually customized their workspaces.}
    \label{fig:snapshot}
    \Description{Snapshots of participants using PromptCanvas during the lab study.}
\end{figure*}

In this section, we present the results of both the lab and field studies conducted to evaluate PromptCanvas. As a first overview, \cref{fig:snapshot} shows one screenshot per participant using PromptCanvas in the lab study.

\subsection{How Do Users Perceive and Interact with the Widgets in PromptCanvas? (RQ1)} \label{sec:results_widgets}
\subsubsection{Perception on the Number of Generated Widgets}
During the tasks in the lab study, participants, on average, created 8 widgets for email writing and 13.5 widgets for short story writing. Regarding the generated number of widgets, participants had mixed opinions. Some (P1, P2, P3) found the number of widgets to be good, with P11 also appreciating the balance, noting, \textit{``It had a good balance of not being too overwhelming and also not being too lacking.''} Some (P4, P5, P6, P8) thought the number of widgets was alright. With P8 stating, \textit{``I think, since we can create more widgets by double-clicking the canvas, this is very standard to keep the four suggested widgets''}. Some participants (P7, P12) suggested that having fewer widgets might be better. P7 indicated that generating four widgets at once was cumbersome with a trackpad, \textit{``If the number would be lower like two with more accuracy, it would have been a bit of more comfort because then I would not require to swipe.''} Whereas, P13 expressed a desire for slightly more, stating, \textit{``I would have expected a bit more, maybe two or three more.''} P14 felt that four widgets were adequate but appreciated receiving more suggestions, saying, \textit{``The number four is good enough. But as it gave more suggestions, it helped me more''}. P17 commented that the number of widgets might depend on the user's experience level. They also appreciated the option to delete or create more widgets, \textit{``You have the option to delete the widgets that you don't need, and you could also create more widgets if you need it''} --P17. P18 suggested, \textit{``I think the number is a good starting point. I don't know if you can make it user-selectable. This would be even more flexible.''}

\subsubsection{Perception of Suggested Widgets}
Participants expressed various opinions about the suggested widgets generated by the system after the lab study. Most of them appreciated the suggestions. The suggestions were useful for both email writing (P2, P3) and story writing (P6, P8, P10, P15). P3 found the widgets helpful for their task on a recommendation letter, noting, \textit{``The widgets seem to be more specific instead of being general. For the recommendation letter, those specific widgets really helped''}. P5 thought the widgets were good, especially appreciating their flexibility: \textit{``Another good thing about the widgets was that it wasn't rigid, so you could add your input as well''}. For the story writing task, P6 found the widgets quite helpful, explaining, \textit{``The widgets that were produced by the system gave me a plot twist and suggestions to edit the introductory part''}. P11 was satisfied with how the widgets identified key areas for modification, saying, \textit{``The generated widgets were really good... they extracted the main points ... the story, which can be modified or changed.''} P12 appreciated our tool's ability to provide numerous ideas and points, saying, \textit{``With the dynamic tool (PromptCanvas), I could write more, and more ideas and points were coming to me that I had not planned before.''} P16 added, \textit{``The suggestions were very accurate''}. P18 elaborated, \textit{``I think they (widgets) got the essence and the few things that may not be immediately apparent''}. P14 valued the extra suggestions that helped them focus on creativity, stating, \textit{``It produced important suggestions and some extra which I couldn't think of at that time''}. 

In contrast, P4 felt that the generated widgets lacked diversity, mentioning, \textit{``The first widgets that I got were related to general information like the company name and years of experience... it would have been nice if there were solutions like the tone or how the email will end.''} Moreover, P7 suggested that more thematic suggestions could enhance productivity. However, P15 liked having multiple suggestions and found them helpful, particularly for understanding how to use the system as a novice: \textit{``They were good, especially because I didn't yet know the system. That made it easy for me to understand the widgets''}.

\subsubsection{Perception of Suggested Values within Widgets}
In the lab study, participants expressed diverse opinions on the suggested values and options for each widget. P1 found the suggestions particularly effective when creating their own widgets, stating, \textit{``whenever I created widgets of my own, then the suggestion was good. When you mention a topic ... They understood many things from there.''} Conversely, some (P2, P4) felt there was a lack of variety, saying, \textit{``They were helpful, but I also think they were not diverse enough in their nature. They were very similar''}. However, they acknowledged an improvement upon requesting more suggestions: \textit{``When I clicked for more suggestions, then the solutions got really better.''}

Some participants (P5, P6, P7, P10, P14) highly appreciated the variations within the widgets, finding them to be fun (P6), accurate (P7), and helpful (P6, P7, P18). Participants (P5, P14) highlighted the ability to choose from multiple suggestions, \textit{``I had fun where even inside the widget suggestions you had options to choose from ... that's great''} --P5. P14 stated\textit{``It really helped me a lot. I really liked that part when the widgets themselves had another suggestion.''} P15 appreciated the flexibility to edit suggestions if needed, whereas P16 had a mixed experience, finding some suggestions very close to what they had in mind, while others were not useful. Furthermore, P18 considered the suggestions \textit{``I think it was very, very useful.''}

Participants further praised the tool's creative suggestions (P5, P11, P6), suggestions being practical (P17), and time-saving (P5, P11). P11 added, \textit{``It presented me with ideas or scenarios which I might not have thought of before.''} P6 appreciated how the tool provided new ideas and allowed for comparison without cognitive effort: \textit{``I don't need to think on my own, I can look there and compare.''} P8 linked the suggestions to having a creative partner, stating, \textit{``Getting suggestions feels like a friend is with me... it felt like this is what we expect from AI.''} P13, while not using the suggestions extensively, found the initial ones helpful as a starting point: \textit{``When I started typing or writing on the, let's say, or replacing the value on the widgets itself, the suggestions that were provided seemed okay for me.''} They added, \textit{``I'm not a professional writer, so I'd like to have some starting point, then I can keep up pace afterward.''} P17 believed the suggestions were well-aligned with the text or direction provided, saying, \textit{``From my perspective, it's good enough... it's wonderful.''}

\subsubsection{Flexibility of Rearranging Widgets}
Participants had varied opinions on the ability to rearrange widgets on the canvas during the lab study. Several (P2, P3, P6, P8, P11, P13, P14, P15, P17) found the flexibility of moving widgets beneficial. P2 found the rearrangement process effortless and easy to use. P3 valued the spatial arrangement, noting, \textit{``The widget that controls the address of the company, I just put it beside where the address of the company actually is. So to maintain that spatial relationship ... the arranging of the widgets really helped''}. P8 found PromptCanvas helpful, comparing it to a canvas that allows for creative expression and breaks the monotony in comparison with the conversational UI: \textit{``This UI (PromptCanvas) is like a canvas where we can paint... it cuts through the boredom''}. Participants, like P4, did not find this flexibility particularly important, while P5 described the experience as \textit{``so good''}.  However, some participants emphasized the value of a fixed placement of widgets, suggesting that color-coding (P12) or a more rigid arrangement (P1, P10) might have been better for their needs. P6 felt the rearrangement feature was helpful, though they needed more time to adapt. They noted, \textit{``If you use it for a long time, and you get used to it, I think it's more fun, and it will help you''}.

P18 suggested that an auto-resizing or auto-fit option might be beneficial for managing a large number of widgets: \textit{``If I have more than like 10, 12 widgets, that can be a bit cumbersome to move around.''} Other participants (P11, P13) appreciated the ability to drag widgets and keep the canvas organized, saying, \textit{``I liked the widgets floating around and also the ... main widget where I was writing and the text was generated. I was free to arrange it in whatever way I liked''} --P11. P13 explained that it allowed them to cluster related widgets together: \textit{``If I have the flexibility to rearrange them anywhere I want on the canvas, it also helps me to cluster them together''}. P15 liked the option to move and resize widgets, which helped manage visual clutter: \textit{``I liked being able to move things around and make them bigger or smaller as I wanted to.''} P17 found the fluidity of moving widgets an \textit{``amazing feature''} that provided a \textit{``lively feeling''}.

\subsubsection{Perception of Widgets: Section Replacements vs. Abstract Guidance}
Several participants in the lab study appreciated both specific section replacements and abstract guidance (P1, P4, P7, P11, P13, P14, P15, P16), finding that each approach had its own advantages depending on the context. Participants (P3, P4) noted that thematic suggestions were important for creative writing, while more specific widgets were useful for structured tasks, \textit{``so for example, if I'm writing a short story then having thematic suggestions or widgets is important but for example if I'm writing something that's a more structured resignation letter... then more specific widgets are useful''} --P3. 

Some participants (P1, P5, P6) found the text replacement particularly intuitive. P1 appreciated it, stating, \textit{``replacing text was really intuitive for me because I replaced it many times in the text''}. However, others (P8, P12) preferred widgets that provided abstract guidance (e.g., tone, length). P12 found that widgets for changing specific text parts were useful for meeting word limits and making necessary adjustments. P8 favored abstract guidance as it did not suppress creativity, stating, \textit{``I think the guidance is more suitable for me; it doesn’t wash off the whole creativity of the creator or the writer.''} On the other hand, P10 leaned towards abstract guidance generally but acknowledged the usefulness of specific widgets in particular cases.

\subsubsection{Perception of Ownership and Control Over Generated Text}
During the lab study, participants reported varying perceptions of ownership and control over AI-generated text, often influenced by the tools and features available in the interface. P11 noted that the sense of ownership differed depending on the task, observing that editing or polishing existing content felt distinct from generating text \textit{``out of the box.''} P3 also found a stronger sense of ownership in shorter, simpler texts, such as emails, compared to longer forms like stories.

Participants felt the widgets enhanced their control over the generated text, stating, \textit{``You have control in the static UI, but it can be cumbersome. It (PromptCanvas) was a more streamlined control, which was really nice''} --P3. One of the participants elaborated on the comprehensive nature of the widgets, noting, \textit{``These widgets are basically suggestions ... and then there are suggestions inside of suggestions which makes it even more comprehensive ... more creative, so many frontiers that you could take ... fantastic tool, it's a great experience to use it''} --P5. P8 noted that widgets help in narrowing down the focus, \textit{``It really narrows down the focus where I need to edit something, so if I click on the widget and get some suggestions, I think I will also be able to select my focal points where I need to change something, where I need to add something more, so that's what I think about the suggestions, that's what I think about all the prompts or all the words that are found in the widgets''}. P10 liked the flexibility of customizing the widgets. Widgets played a key role in providing a \textit{``greater sense of control''} — P3. For example, P15 described feeling a sense of \textit{``creative control''} when using widgets, stating, \textit{``I felt like it was my story because the widgets weren’t writing the entire text all at once, but rather in small pieces,''} giving them the impression that they, rather than the system, were deciding what should be written. P15 also appreciated that \textit{``you could go back and edit a previous widget without it changing the entire text.''}

Although PromptCanvas emphasizes widgets, \textbf{participants expressed a strong preference for being able to manually edit the text in the editor}. P15 mentioned, \textit{``I like that you can just go into the text and edit it yourself and add a sentence.''}
P11 emphasized efficiency, stating, \textit{``More control in less amount of time''}. P12 compared PromptCanvas favorably to the conversational UI baseline (referred to as ``static''), noting that the latter often cut off previous text: \textit{``I think when I was regenerating the text in static, I noticed that some of my previous texts were being cut off and I did not like that. And I think that in that case, the dynamic interface was a lot more helpful. And it included the previous ones ... that I was trying to add, and the new ones. So the summarized newer text was what I actually wanted to write, not that it was cutting off something. I find it really important''}. Moreover, P15 found the widgets intuitive and beneficial for beginners. However, P16 found the new widgets required more learning but acknowledged their advantages. 

\subsubsection{Perception after Prolonged Use in the Wild} 
\label{subsec:studywild}
In the interview following the two-week field study, all participants maintained a positive opinion of the system, finding the extended usage beneficial. When asked if the additional time was necessary for understanding the system or if the initial lab study time was sufficient, nine out of ten participants agreed that the extended time was not required, as they \textit{``already had good understanding''} --P11. P3 mentioned that the core functionality was straightforward and \textit{``with regards to how to use it, I don’t think we need the extended time.''} Other participants shared similar views, stating \textit{``it’s good having more time with it, so I could experiment and try different things''} --P15, and \textit{"that I had more time to explore and more time to play around. So that's what actually having no time limit benefited me"} --P11. In contrast, P4 noted that in the lab study \textit{``there was some time constraint and the system was new, so then it felt like the system was a bit difficult to understand, but when I had much time, then the system, the overall environment, became very intuitive to me and after some time, I had no issues in conducting the whole setup. So I think time helped a lot and the more I used the system, the more I started to like it.''}

Over the two weeks, participants tested the system across various use cases. P2 engaged in writing motivational letters, book reviews, and stories. P3 focused on email replies and creative writing. P17 used it for critical analyses, while P11 explored prompt generation for image creation. P11, P13, and P14 tested programming tasks, with P14 remarking, \textit{``It really performed well.''} P14 also generated search engine optimized content and tested the system’s ability to work with other languages by using Bengali content. Additionally, several participants expressed that image support would be both beneficial and an \textit{``intuitive need''} --P17. P11 tested this by generating prompts within PromptCanvas and then using them in an external text-to-image generator, reporting positive outcomes. With extended use, participants became more effective in using the system. For example, P15 achieved better results using multiple widgets for small, modular aspects instead of high-level instructions. They adopted a workflow using widgets for modifications and prompts for editorial decisions, such as introducing a new character with a widget and later moving its introduction within the story.

\subsection{How Does PromptCanvas Support Creativity and Exploration? (RQ2)}
\label{subsec:creativity}
PromptCanvas scored significantly higher than the baseline on all factors of the Creativity Support Index (CSI) (all $p < 0.03$, see \autoref{tab:CSI-within}) in our lab study. As our study did not involve collaboration, we omitted the collaboration factor following the practice from~\cite{carollCreativityFactorEvaluation2009, codetoon} to avoid confusion. Participants found PromptCanvas ($M=82.09, SD=12.12$) to support creativity significantly more ($p=0.005$) compared to the conversational UI ($M=61.65, SD=18.53$). The $p$-values were adjusted using the Bonferroni-Holm correction to account for multiple comparisons. The Creativity Support Index results from our two-week-long field study, shown in \autoref{tab:CSI-field}, found similar results ($M=79.73, SD=17.73$).

\begin{table}[t]
\caption{Creativity Support Index (CSI) results from the lab study (N=18).}
\begin{tabularx}{\linewidth}{Xd{2.2}d{2.2}d{2.2}d{2.2}d{0.3}}
\toprule
   &  \multicolumn{2}{c}{Baseline} & \multicolumn{2}{c}{PromptCanvas} & \\ %
   \cmidrule(r){2-3}\cmidrule(lr){4-5}%
   Factor  & \multicolumn{1}{c}{M}  & \multicolumn{1}{c}{SD} & \multicolumn{1}{c}{M}  & \multicolumn{1}{c}{SD} & \multicolumn{1}{c}{$p$}\\ 
   \midrule
    Enjoyment& 13.06 & 4.40 & 16.56 & 3.99 & .02\\
 Exploration& 11.78 & 5.53 & 16.83 & 3.08 &  .02\\
 Expressiveness& 10.83 & 4.91 & 14.67 & 3.76 & .02\\
 Immersion& 10.00 & 4.10 & 14.61 & 4.73 & .01\\
 Results Worth Effort& 14.17 & 4.02  & 17.61 & 2.21 & .005\\
    \midrule
    Overall CSI Score& 61.65 & 18.53 & 82.09 & 12.12 &  .005 \\ 
\bottomrule
\end{tabularx}
\Description{This table shows the Creativity Support Index (CSI) results from the lab study (N=18).}
\label{tab:CSI-within}
\end{table}

\begin{table}[ht!]
\caption[Creativity Support Index results from our field study.]{Creativity Support Index (CSI) results from our field study (N=10).}
    \centering
    \begin{tabular}{lcc}
    \toprule
          Factor&Avg. Score & SD\\
    \midrule
 Enjoyment& 16.50 & 3.58\\
 Exploration& 17.20 & 2.48\\
 Expressiveness& 15.10 & 4.85\\
 Immersion& 12.90 & 4.25\\
 Results Worth Effort& 16.40 & 3.75\\
 \midrule
 Overall CSI Score& 79.73 & 17.37 \\
\bottomrule
    \end{tabular}
    \Description{This table shows the Creativity Support Index (CSI) Results from the field study (N=10).}
    \label{tab:CSI-field}
\end{table}

\begin{figure*}[t]
    \includegraphics[width=1\textwidth]{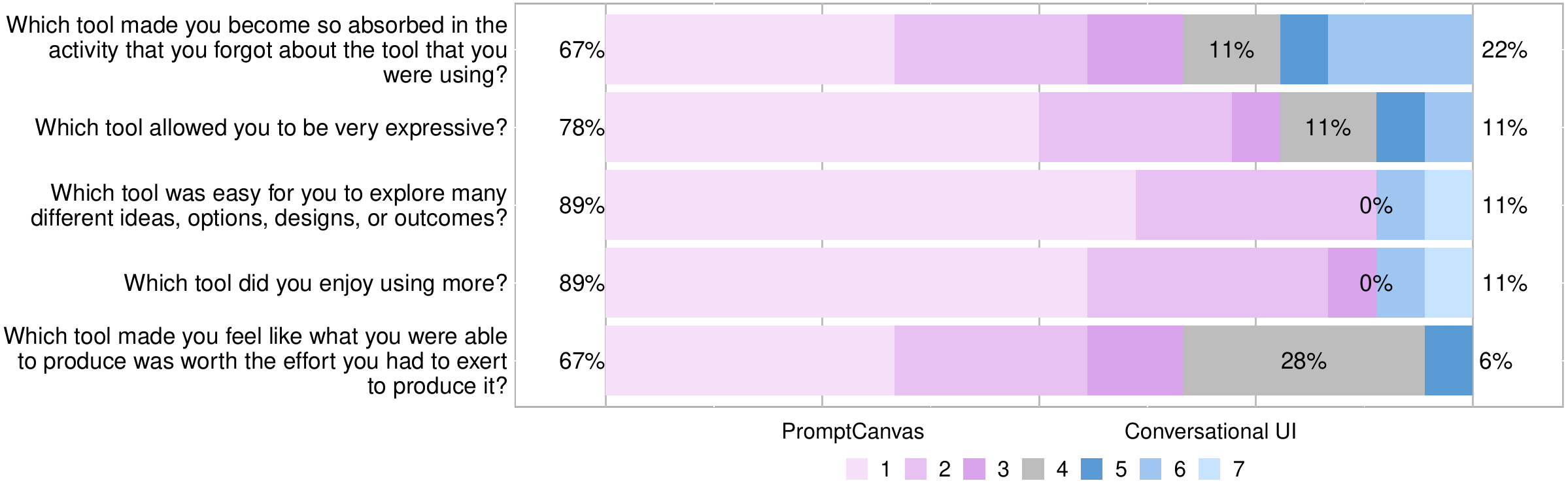}
    \caption{Self-reported creativity support scores and preferences comparing PromptCanvas and the Conversational UI (N=18).}
    \label{fig:Preference_CSI}
    \Description{This image shows participants' self-reported creativity support scores comparing the baseline and our system.}
\end{figure*}

In the final survey of the lab study, participants directly compared the creativity support between PromptCanvas and the conversational UI (\autoref{fig:Preference_CSI}). 67\% of participants reported that they became so absorbed in the activity that they forgot about the tool they were using. 78\% of participants chose PromptCanvas as the more expressive tool. 89\% of participants found that they explored a wider range of ideas, options, designs, or outcomes using our system compared to using the baseline.
Furthermore, 89\% of participants reported a higher level of enjoyment with PromptCanvas than with the baseline. Using PromptCanvas made users feel their efforts were most worthwhile, with 67\% feeling satisfied with what they produced relative to the effort expended.

\subsubsection{Exploration}
After the lab study, all participants generally felt that the widgets helped them in finding unexplored areas or ideas. P14 mentioned, \textit{``For example, when I was writing about the story of AI robots, I was only thinking about a dystopian future where AI controls society. But when I got the suggested widgets, I was able to add the AI's influences on society specifically when AI will interfere with human affairs and national politics. It was not given at first in the generated story, but I could add it then. I couldn't think of it if the widgets didn't suggest this. So it was really helpful. It made the story more suspenseful and good.''} P8 added, \textit{``... when these dynamic widgets are suggesting something, it is also giving us the ideas to explore more in those areas, so I think this is really cool''}. Participants (P10, P13) also appreciated how the widgets allowed them to incorporate new ideas and details into their writing without reverting the text to its initial state.

Participants found PromptCanvas helpful for creative writing (P1, P2, P14). They noted that the tool encouraged them to think in multiple directions (P1, P2), \textit{``I would say that it helped me to think more creative things, but when I was thinking, it helped me to think in multiple ways''} --P1. In addition, P3 added, \textit{``If it's a more open-ended task ... you don't really know what ... the final form of your creative writing might be. Then I think those widgets would help ... having suggestions for more general themes, especially for the short story portion, would definitely help a lot''}. Furthermore, The dynamic nature of widgets was described as a ``game changer'' by P5, who valued the ability to generate multiple lines of thought quickly. This aspect allowed for rapid exploration of different story directions, \textit{``Dynamic in nature, that's a game changer, as I said, because those lines of thought, I mean, I'm not saying that they wouldn't pop up, but it'll take a whole lot of time for me to... bring those up manually''}.

\subsubsection{Diversity}
Participants mentioned that widgets help avert the monotony of the generated text by making it more personalized to the individuals (P4, P12). They appreciated the variety of suggestions provided, which improved the quality and helped bring out the personal styles of their writing, \textit{``When you are writing something, sometimes we are just too absorbed in the generated result that we are not thinking about the possibilities in which we can improve our writing more, so having widgets that will suggest us more options can really improve the writings''} --P4. P4 also noted, \textit{``I think in making people's writing more personal to them, they (widgets) just can be really helpful, the writings won't be monotonous''}.

Participants found PromptCanvas supports both structured input (P6) and diversity (P7, P10). P7 and P15 further mentioned the advantage of having visible prompts through widgets, facilitating continuous editing and refinement of the text, \textit{``What happens in static UI is that you continuously are giving prompts... so once a new prompt is given, and it has been applied the prompt goes out of the window. So when I'm having the widget, I can see what are my exact prompts that are functioning''} --P7. P15 valued the ability to see an overview of the story through widgets, which allowed for focused changes: \textit{``It just comes down to seeing the overview of the story physically represented by the widgets. It felt like, again, that's how my mind pictures things. I like breaking things up into pieces. And so having the widgets, seeing that part of the story broken up means I could focus down on that and change it without changing the entire story. And then adding that into the text. And it only changed that part of the text''}. P15 added, \textit{``... the widgets, having it broken down in front of me and seeing the individual things. So that would be a bit more creative, I think''}. Participants (P9, P12) also mentioned that widgets helped them think outside the box, offering diverse points and even aiding in non-creative tasks like writing biographies.

\subsection{How Does PromptCanvas Affect the Cognitive Load in Creative Writing? (RQ3)}\label{subsec:cognitive_load}

\subsubsection{NASA-TLX} 
\begin{figure*}[t]
    \centering
    \includegraphics[width=\linewidth]{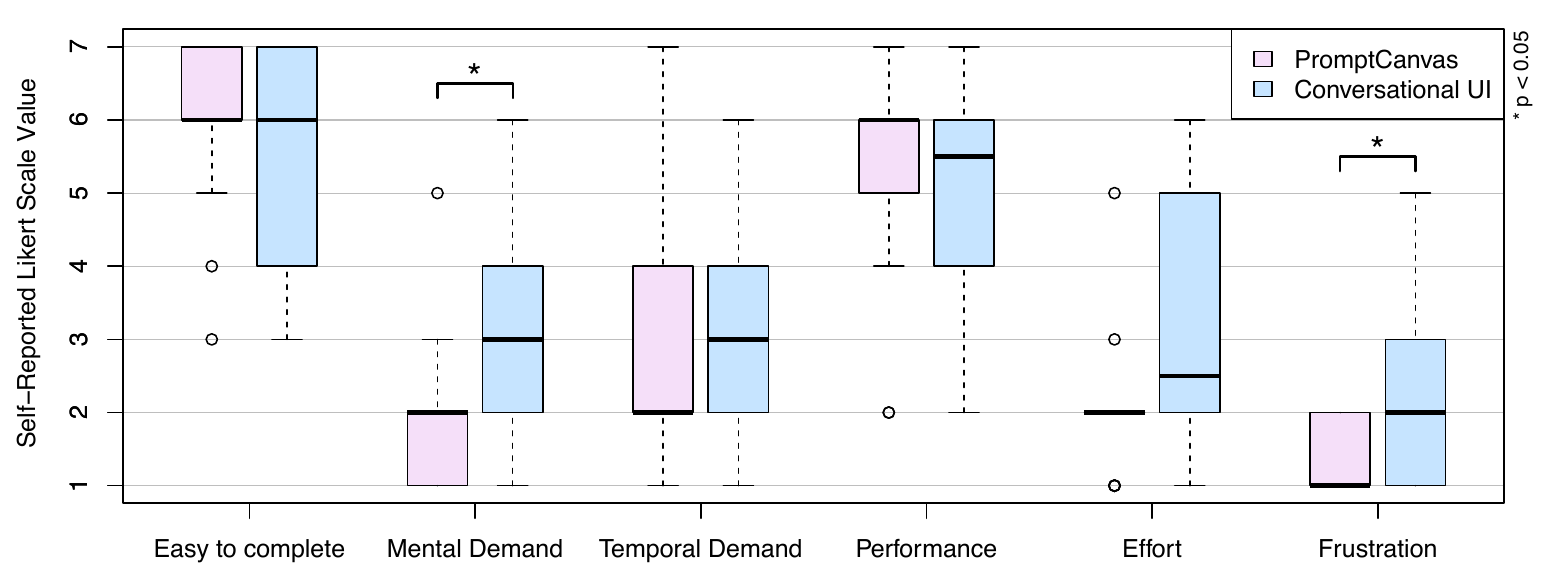}
    \caption{Self-reported NASA-TLX scores and ease-of-use ratings from participants in our lab study (N=18).}
    \label{fig:NASA-within}
    \Description{This box plot shows participants' self-reported scores for NASA TLX questions and ease of using the system.}
\end{figure*}

Participants evaluated their perceived cognitive load using the NASA-TLX scale after using each UI in the lab study. Significant differences were observed in two aspects of cognitive load: mental demand and frustration. For mental demand, participants reported a significantly ($p=0.02$) lower mental demand when using the dynamic interface ($M=1.89$, $SD=1.02$, $Med=2$), compared to the static interface ($M=3.06$, $SD=1.51$, $Med=3$) (See \autoref{fig:NASA-within}). Frustration ratings also differed significantly ($p=0.03$), with the dynamic interface showing lower frustration ($M=1.28$, $SD=0.46$, $Med=1$) than the static interface ($M=2.17$, $SD=1.42$, $Med=2$), indicating that PromptCanvas helped reduce feelings of insecurity, irritation, and stress. However, other cognitive load aspects showed no significant differences between the two interfaces.

\begin{figure*}[t]
    \centering
    \includegraphics[width=\linewidth]{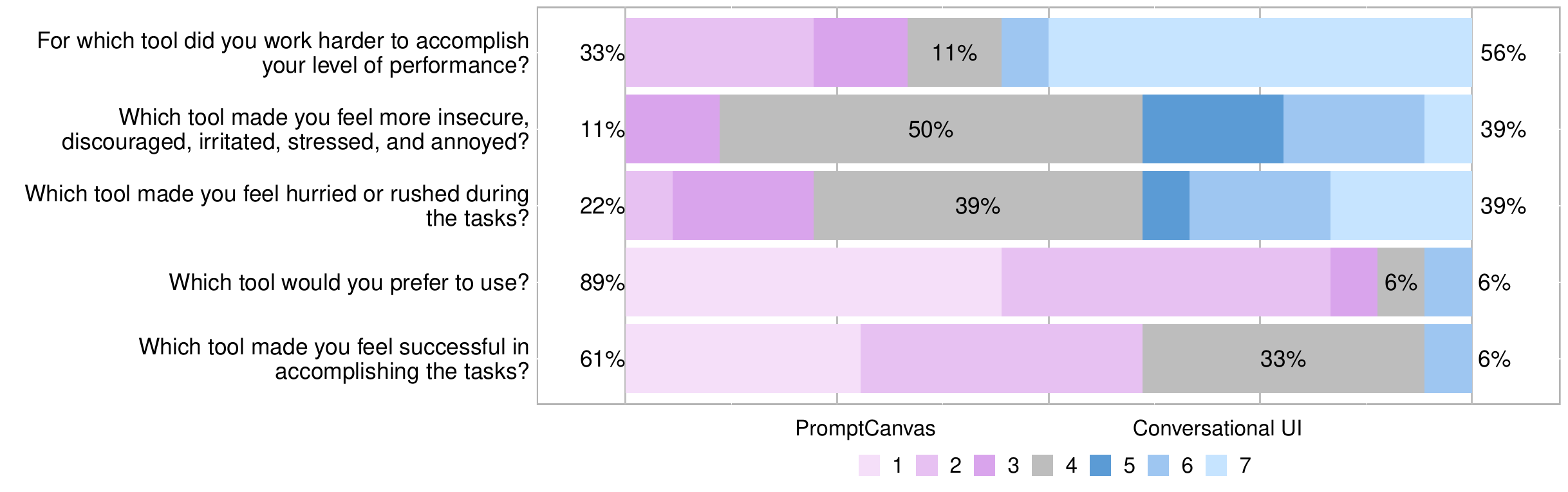}
    \caption{Self-reported cognitive load and preference scores comparing PromptCanvas and the conversational UI (N=18).}
    \label{fig:Preference_NASA}
    \Description{This image shows participants' self-reported cognitive load and preference scores that directly compare the Conversational UI and PromptCanvas.}
\end{figure*}

In the final survey of the lab study, participants directly compared their perceived cognitive load between PromptCanvas and the conversational UI. Results are shown in \autoref{fig:Preference_NASA}. These results show that when using PromptCanvas, 39\% of our participants felt less frustrated, while 50\% did not feel frustrated with either UI. Additionally, 39\% perceived less temporal demand with PromptCanvas, while 39\% of participants did not feel hurried or rushed with either UI. Regarding the feeling of success, 61\% of participants felt more successful in accomplishing tasks when using PromptCanvas, and lastly, 56\% of participants reported needing less effort with PromptCanvas to accomplish their level of performance.

\subsubsection{Widget Complexity}
During the field study interviews, participants were asked if using widgets justified the additional complexity they introduced. P3 pointed out that the usefulness of widgets depended on the task's length: for short responses, they seemed unnecessary, but for longer or more creative outputs, like creative writing or full emails, they provided valuable control. Several participants appreciated how widgets reduced the need for multiple prompts. P11 preferred creating multiple widgets over composing several prompts. At the same time, P18 found them \textit{``definitely worth it,''} as they eliminated the need to \textit{``write multiple prompts or go through this chain of changing prompts.''} P18 also mentioned that the widgets were useful as they visually preserved the context. P17 did not find the widgets complex, describing them as \textit{``fantastic and very logical,''} with \textit{``short sentences or short, relevant words.''} However, P3 acknowledged that \textit{``using widgets does take a bit more time.''} P8 felt this time investment was worthwhile for creative tasks, saying, \textit{``a bit time-consuming, however, I am actually putting that time to bring out my creativity and bring out new ideas, so if I’m focusing on writing.''}

While some participants noted that widgets added complexity, most viewed this as a beneficial trade-off, especially for tasks requiring detailed control and creativity.

\subsubsection{Efficiency and Efficacy}
In the lab study, participants using PromptCanvas provided significantly ($p=0.0006$) fewer prompts ($M=4.00, SD=2.68$), compared to the conversational UI ($M=11.11, SD=6.65$). P14 highlighted the efficiency of PromptCanvas, noting that \textit{``if I had to change anything in the generated text in the static UI, I have to give a comment again and again and check that out. But in the dynamic UI, it was quite efficient. I didn't have to wait for the full generation like this. And I can only give comments to a specific area of the generated text in the widgets. It was quite helpful.''} Similarly, P17 appreciated the flexibility provided by PromptCanvas, saying \textit{``The static one was more in one direction, so with the dynamic one (PromptCanvas), it was really like, okay, I don't have to push myself to you know harder to explore things or to create something and get the result.''} Additionally, P18 valued the ease of directing the tool with less manual intervention, stating \textit{``I think that the widgets helped me to direct the tool to a more specific ... direction that I wanted to go without having to manually intervene as much. I think that we can do similar things with the static UI, but then I have to write a lot more prompts.''} P18 further added, \textit{``So, even though the end result might be very similar, the efforts of getting there seems like it is definitely more with the static UI.''}

\subsection{System Usability}
After the two-week field study, participants generally reported positive feedback towards the system's usability. P17 complemented the interface as \textit{``very good''} and \textit{``eye soothing''}, while P15 mentioned they \textit{``like the colors ... It's easy to look at for a long time.''} Further, P15 found the widgets intuitive and beneficial for beginners, while P16 found the new widgets required more learning but acknowledged their advantages. 

\subsubsection{System Usability Scale}
\label{sec:ResultsSUS}

\begin{figure}[ht!]
    \centering
    \includegraphics[width=\linewidth]{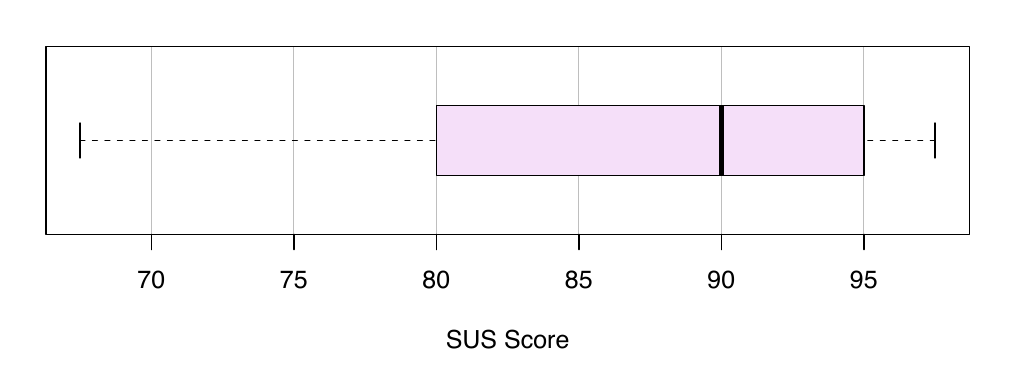}
    \caption[System Usability Scale (SUS) results from our field-study.]{System Usability Scale results ($M = 86.50, SD = 8.96$) from our field-study (N=10).}
    \label{fig:SUS}
    \Description{This image shows the System Usability Score for the second version of our system.}
\end{figure}

After having engaged in both studies, all 10 participants in the field study completed the SUS questionnaire, which yielded an average score of 86.50 ($SD = 8.96$), with a median of 90.00 and a range from 67.50 to 97.50, visualized in \autoref{fig:SUS}. These results rank PromptCanvas well above the usability benchmark of 68, with most scores falling in the “excellent” range, typically above 85~\cite{sus_eval}. The middle 50\% of users rated the system between 80.63 and 93.75, underscoring strong satisfaction with the interface’s usability. The minimum score of 67.50 is close to the acceptable threshold, while the maximum of 97.50 shows that some users found the system nearly flawless.

\subsection{Which Tool Do Users Prefer (PromptCanvas Vs. Conversational UI)?}
The results of the lab study indicate a strong preference for PromptCanvas compared to the baseline conversational UI.
\subsubsection{General Preference}
Notably, 89\% preferred PromptCanvas over the conversational UI (\cref{fig:Preference_NASA}). P1 recognized our tool as a canvas, highlighting its strengths in exploration and creativity. They suggested that while the conversational UI is suitable for time-constrained tasks, PromptCanvas might excel in scenarios where there is more time to explore options. \textit{``I would suggest one thing. If you give a time test, I think this static UI is good. But if you don't have any time constraints, then dynamic UI is very good. Because you are exploring many things''} --P1. Similarly, P18 appreciated PromptCanvas’s flexibility in selecting different story settings and character details, finding it beneficial for regular writing tasks: \textit{``This is, I think, something that I would appreciate in regular use.''}

\subsubsection{Ease of Use}
Participants also found it easier to complete tasks using PromptCanvas than the conversational UI (\cref{fig:NASA-within}). P4 found PromptCanvas particularly convenient, noting that it allowed for easy changes with simple clicks, contrasting this with the more cumbersome process in the conversational UI (baseline): \textit{``The dynamic UI (PromptCanvas) was a very convenient tool. In a static UI (Conversational UI), I have to explain to the tool every time I need a change, but in a dynamic UI, I can achieve that with a simple click.''} P10 appreciated the increased editing opportunities provided by PromptCanvas: \textit{``It provides you with more possibilities to do.''} P2 enjoyed the spontaneity of modifications without needing detailed instructions. Additionally, P12 found PromptCanvas versatile and engaging compared to the monotonous Conversational UI: \textit{``The static tool was a bit monotonous ... I thought it had way fewer options, and the dynamic one was really versatile and not boring.''}

\section{Discussion and Future Work}
In this work, we investigated how dynamic widgets can be used to improve user interaction and creativity and provide greater control over generated content \textbf{(RQ1)}. We also assessed whether dynamic widgets for iterative and structured prompting offer better creativity support compared to a conversational UI \textbf{(RQ2)}, and explored whether they help reduce cognitive load in creative writing tasks \textbf{(RQ3)}. The results indicate that PromptCanvas enables a more exploratory and iterative approach, allowing users to experiment with different aspects of their work more freely. Below, we will address some of the current limitations while emphasizing opportunities for future research.

\subsection{Study Limitations}
The studies conducted with PromptCanvas have several limitations regarding the sample size and generalizability of the findings. The lab study involved 18 participants, while the field study included 10, which may not fully capture the diversity of user experiences in broader populations. Future research could include a larger, more varied sample and examine how professional writers use PromptCanvas. 

\subsection{Widget Organization and Interdependence}
Although multiple participants found our UI to be aesthetically pleasing, participants encountered issues with organizing widgets on the canvas (see \cref{sec:results_widgets}). As a solution, having widgets clustered based on their purpose could enhance usability. This organization would allow users to more easily navigate the interface and locate the tools they need for specific tasks. Moreover, implementing a color-coding system for widgets was another suggestion. 
This might help users quickly differentiate between various widgets and their functions, improving the overall usability of PromptCanvas. Participants also suggested that it would be helpful to highlight sections or changes related to specific widgets. This feature could improve the user experience by making it easier to track and manage modifications. Finally, widget organization by the user is currently unused by the system but could actually serve as an additional input channel to the AI, conveying, for example, likely relationships between concepts for which the user has grouped the widgets closely together.

The widgets also currently form self-contained interface objects that are not coupled to other widgets. However, changes in one widget might, on a semantic (as opposed to structural) level, have influences on other widgets, which in turn would need to change their values or available options. As an example, assume that the princess in the story suddenly has a son instead of a daughter, which would most likely influence their proposed names. A future iteration of our concept could support this.

\subsection{Supporting Creativity and Exploration}
PromptCanvas significantly supports creativity compared to the conversational UI (\textbf{RQ2}). The results from \cref{subsec:creativity} demonstrate that users find PromptCanvas more helpful in creative writing and idea generation. These results also suggest that most users explored themes they might not have considered without the dynamic widget suggestions provided by PromptCanvas, similar to observations by \citet{dynavis_2024} and \citet{copilot}. Unlike the static nature of traditional conversational UIs, PromptCanvas offers a dynamic, interactive environment where users can manipulate various widgets to explore and refine their creative outputs. This approach not only enables a more engaging creative process but also supports non-professional writers by providing intuitive tools that help them overcome obstacles in creative writing. By enabling users to experiment with different text elements and receive contextually relevant suggestions, PromptCanvas empowers individuals with less experience in creative writing to generate and express their ideas more effectively (\cref{subsec:cognitive_load}).

\subsection{Supporting Customizability and Control}
Our study suggests that PromptCanvas offers advantages in personalization and customizability (\cref{sec:results_widgets}) through its dynamic widgets (\textbf{RQ1}). These widgets allow users to customize their interface to better suit individual preferences and needs. Users also reported feeling a greater sense of control over the generated text (\cref{sec:results_widgets}), which aligns with the principles of managing one's cognitive resources~\cite{metacognition}. According to \citet{metacognition}, metacognition is the psychological ability to monitor and regulate one's thoughts and behaviors, and generative AI systems place metacognitive demands on users. We envision that the dynamic widgets of our generative AI frontend, PromptCanvas, can reduce metacognitive demand by enhancing customizability. Customizability can support cognitive control by providing a more adaptable interface~\cite{KHAMAJ2024164}. 

Moreover, current generative AI systems often require verbalized prompting, demanding self-awareness of task goals and decomposition of tasks into sub-tasks~\cite{metacognition}. The results from \cref{subsec:creativity} show that participants viewed dynamic widgets as a means to break down tasks, enhancing their ability to manage and execute complex workflows. Additionally, users found that using PromptCanvas made the results worth the effort they put in (\cref{subsec:creativity}). By guiding reflection and prompting towards a more structured and interactive use of generative AI, our concept of PromptCanvas might be able to help users engage more deeply with their creative tasks.

Results from \cref{sec:results_widgets} indicate that, in the field study, participants used PromptCanvas not only for creative tasks but also for programming, demonstrating the system's versatility beyond creative writing. This extension into non-writing tasks further emphasizes the customizability and adaptability of PromptCanvas, allowing users to personalize their experience for a variety of workflows. This broader applicability highlights the system's potential to serve diverse needs, extending beyond writing to other domains such as coding.

\subsection{Reducing Cognitive Load}
Although the dynamic nature of PromptCanvas could be expected to increase cognitive load~\cite{dynavis_2024}, the results from \cref{subsec:cognitive_load} show that the NASA-TLX ratings revealed statistically significant differences in mental demand (\textbf{RQ3}) and frustration between the UIs. Participants noted feeling less annoyed, more productive, and able to complete tasks with less effort, suggesting that the dynamic interface offers a more engaging and efficient creative process. By allowing users to manipulate widgets for contextual suggestions and seamless exploration, PromptCanvas creates a supportive environment for managing creative writing tasks.

\subsection{Extending the Concept to Other Domains}
Inspired by previous research in visualization (\textit{DynaVis}~\cite{dynavis_2024}) and systems like \textit{Luminate}~\cite{suh2024luminate}, we expect the concept of PromptCanvas to be applicable more broadly. For instance, results from \cref{sec:results_widgets} suggest that, it could extend beyond text-based creative tasks to visual content generation, where dynamic widgets might allow users to iteratively refine and customize image outputs. This capability would enable users to explore artistic styles, integrate specific elements, and adjust parameters, showcasing the versatility of dynamic widgets in diverse creative domains. The canvas-based design of PromptCanvas also allows for a broad range of widget types. Future additions could include more standard HTML elements like date-pickers, sliders, and checkboxes, similar to those in \textit{DynaVis}~\cite{dynavis_2024}, or specialized tool sets from systems like \textit{Textoshop}~\cite{massonTextoshopInteractionsInspired2024}, to allow for more targeted customization and user control.

\section{Conclusion}
In this work, we introduce PromptCanvas, using dynamic widgets as a novel solution to address the limitations of current UIs for generative AI in creative writing. Our studies demonstrate that by incorporating customizable interactive elements, our system enhances user control, reduces cognitive load, and supports iterative exploration and the creation of a personalized workspace. The findings reveal that dynamic widgets significantly improve user experience and facilitate more effective, user-centered interaction with AI. This research emphasizes the importance of user-driven customization and flexibility in unlocking the creative potential of humans in AI-driven writing support, leading to more meaningful and productive writing interactions.

\begin{acks}
Funded by Elitenetzwerk Bayern and the Deutsche Forschungsgemeinschaft (DFG, German Research Foundation) -- 525037874.
\end{acks}

\bibliographystyle{ACM-Reference-Format}
\bibliography{bibliography}

\end{document}